\documentclass[prb,showpacs,footnoteinbib,twocolumn,unsortedaddress]{revtex4}
\usepackage{amsmath,amssymb,epic,eepic}
\usepackage[squaren]{SIunits}
\usepackage[latin1]{inputenc}
\usepackage{graphicx}
\usepackage{tikz}		
\usepackage{circuitikz}
\usetikzlibrary{%
	shapes.geometric,calc,intersections,%
	decorations.markings,decorations.text,decorations,%
	patterns,decorations.pathmorphing,fpu,%
	positioning,automata,shapes.multipart,circuits,%
	circuits.ee, circuits.ee.IEC,decorations.pathreplacing
}

\begin{document}

\title{Real time decoherence of Landau and Levitov quasi-particles in quantum Hall edge channels}

\author{D. Ferraro$^{1,2,3}$}
\author{B. Roussel$^1$}
\author{C. Cabart$^1$} 
\author{E. Thibierge$^1$}
\author{G. F\`eve$^4$}
\author{Ch. Grenier$^5$}
\author{P. Degiovanni$^1$}

\affiliation{(1) Universit\'e de Lyon, F\'ed\'eration de Physique A.-M. Amp\`ere,\\
CNRS - Laboratoire de Physique de l'Ecole Normale Sup\'erieure de Lyon,\\
46 All\'ee d'Italie, 69364 Lyon Cedex 07, France}

\affiliation{(2) Aix Marseille Université, CNRS, CPT, UMR 7332, 13288 Marseille, France} 

\affiliation{(3) Université de Toulon, CNRS, CPT, UMR 7332, 83957 La Garde, France}

\affiliation{(4) Laboratoire Pierre Aigrain, Ecole Normale Sup\'erieure, \\ 
CNRS, Université Pierre et Marie Curie,  Université Denis Diderot,\\
24 rue Lhomond, 75231 Paris Cedex 05, France}

\affiliation{(5) Institute for Quantum Electronics, ETH Zurich, 8093 Zurich, Switzerland}

\begin{abstract}
Quantum Hall edge channels at integer filling factor provide a unique test-bench to understand
decoherence and relaxation of single electronic excitations in a ballistic quantum conductor. In this Letter,
we obtain a full visualization of the decoherence scenario of energy (Landau) and time 
(Levitov) resolved single electron 
excitations at filling factor $\nu=2$.
We show that the Landau excitation exhibits a fast relaxation followed by spin-charge separation whereas
the Levitov excitation only experiences spin-charge separation. We finally suggest to use 
Hong-Ou-Mandel type experiments to probe specific signatures of these different scenarios.
\end{abstract}

\keywords{quantum Hall effect, quantum transport, decoherence}

\pacs{73.23.-b,73.43.-f,71.10.Pm, 73.43.Lp}

\maketitle

The recent demonstration of on-demand single electron sources able to inject
single electronic excitations into quantum 
Hall edge channels~\cite{Feve:2007-1,Leicht:2011-1,Fletcher:2013-1}
or 2DEG~\cite{Dubois:2013-2,Hermelin:2011-1} has opened a new
era of quantum coherent electronics. By combining these sources
to electronic beam splitters, experiments analogous
to the celebrated Hanbury-Brown and Twiss and Hong-Ou-Mandel~\cite{Olkhovskaya:2008-1} experiments 
have been demonstrated at the single electron level~\cite{Bocquillon:2012-1,Bocquillon:2013-1} 
thus opening the way to electron quantum optics~\cite{Degio:2011-1,Bocquillon:2013-3}.
However, this emerging field goes beyond a simple analogy with photon
quantum optics: first of all, the Fermi statistics of electrons differs from the Bose
statistics of photons and leads to the Fermi sea, a
state with no analogue in photon quantum optics. Moreover, electrically charged electrons
experience Coulomb interactions which, despite screening, 
are expected to induce strong decoherence effects as demonstrated by Mach-Zehnder interferometry
experiments~\cite{Ji:2003-1,Roulleau:2007-2,Neder:2007-4,Roulleau:2008-2,Huynh:2012-1}.
Moreover, an in-depth study of
the relaxation of a non-equilibrium electronic distribution at filling factor $\nu=2$ has shown that
a description of the quantum Hall edge channels in terms of Landau quasi-particle excitations
is not valid~\cite{Altimiras:2010-1,LeSueur:2010-1}. 

Although the above results on single electron relaxation and decoherence have
been obtained by considering stationary sources and time averaged quantities such as the
electron distribution function, the recent demonstration of the electronic Hong-Ou-Mandel (HOM)
experiment~\cite{Bocquillon:2013-1} calls for a time resolved approach to 
single electron coherence taking into account the finite duration of the electronic excitation. 
The aim of this Letter is precisely to discuss the
decoherence scenario of two single electron excitations emitted by state of the art sources. First, the
Landau quasi-particles correspond to a Lorentzian wave packet
in energy emitted by a properly operated quantum dot\cite{Feve:2007-1}.
Secondly, the Levitov quasi-particles or Levitons~\cite{Dubois:2013-2} are the minimal single electron 
state obtained by applying
a Lorentzian time-dependent potential with quantized flux~\cite{Levitov:1996-1}. We show that comparing the real time aspects of their decoherence brings
a better understanding of the underlying mechanisms of electronic decoherence whose specific features could be experimentally tested
in HOM experiments.

\medskip

Recently, finite frequency admittance measurements have demonstrated
the existence of collective neutral and charged modes of the 
$\nu=2$ quantum Hall edge channel system~\cite{Bocquillon:2012-2}. 
At low energy, this experiment has 
validated the physical picture of two non dispersive 
eigenmodes emerging from strong inter-channel effective Coulomb interactions: 
the fast mode carries the total charge
whereas the slow one is neutral (dipolar) and therefore, in spin polarized channels, 
is expected to carry only spin. 
The existence of these two modes leads to 
the well known image of spin-charge separation~\cite{Levkivskyi:2008-1} clearly
valid for collective excitations generated by a time dependent voltage 
pulse~\cite{Grenier:2013-1} such as the Levitons~\cite{Dubois:2013-2}. 

But is this simple image still valid when considering an arbitrary single electron excitation? 
In particular, can spin-charge separation be observed for an energy resolved single electron excitation? 
Looking at the time dependent average current, the answer
is certainly yes: at strong inter-channel coupling, the current pulse associated with a single electron excitation is expected to
split into two current pulses propagating at the slow and fast velocities and carrying a 
fractional charge~\cite{Berg:2009-1,Neder:2012-1,Grenier:2013-1}. Recent time resolved measurements of electrical currents
have confirmed this picture for two counter-propagating edge channels~\cite{Kamata:2014-1}.
On the other hand, the energy relaxation does not show such
a splitting between charge and spin: the energy resolved excitation disappears
into a wave of electron/hole excitations~\cite{Degio:2009-1,Degio:2011-1}. 
As we shall see in this Letter, the proper way to reconciliate these two apparently discording
points of view is to work within the Wigner function approach~\cite{Ferraro:2013-1} thus
gaining access to both the time course and energy content of single
electron coherence. 

\medskip

\begin{figure}
\includegraphics[width=8cm]{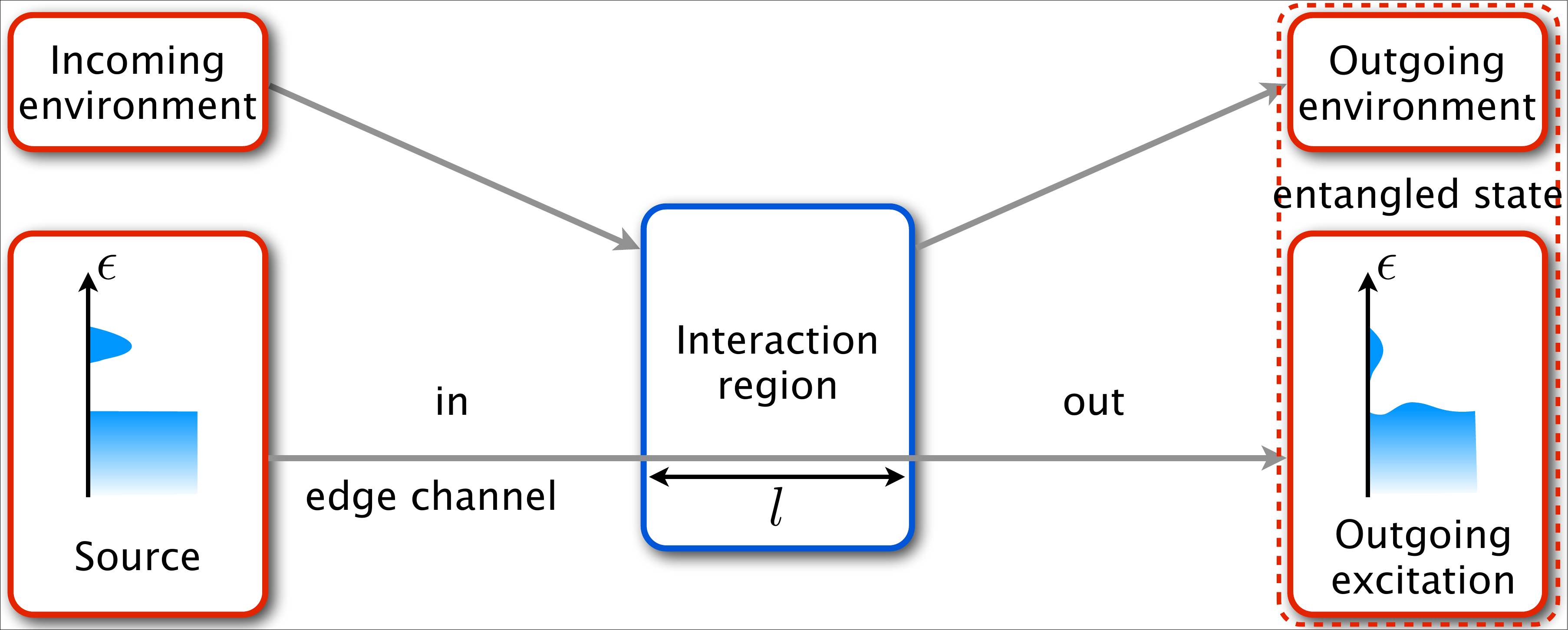}
\caption{\label{fig:input-output} 
A single electron excitation prepared in a coherent wave packet $\varphi_{\mathrm{e}}$ enters
a finite length interaction region. The edge channel comes out entangled with the environmental
degrees of freedom. In the case of two edge channels, the second channel plays the role of 
the environment.}
\end{figure}

To this end, we consider the relaxation of an arbitrary single electron excitation described by a many-body state
$|\varphi_{\mathrm{e}}, F\rangle = \int \varphi_{\mathrm{e}}(x)\psi^\dagger(x)|F\rangle\, \mathrm{d}x$
involving an arbitrary normalized purely electronic wave packet $\varphi_{\mathrm{e}}(x)$ above the Fermi sea $|F\rangle$. 
The quantity of interest is the single electron coherence 
$\mathcal{G}^{(e)}_{\rho,x}(t,t')=\langle \psi^\dagger(x,t')\psi(x,t)\rangle_\rho$
at position $x$ and in the time domain~\cite{Degio:2011-1,Haack:2012-2}. In the present letter, the excitation
is injected into an interaction region as depicted on Fig.~\ref{fig:input-output}.  Our main result is the non-perturbative 
computation of the outcoming single electron conherence ($x$ in the out region) in terms of the incoming 
wave packet $\varphi_e(x)$ and of the interactions. To visualize the result, 
the Wigner function defined as~\cite{Ferraro:2013-1}:
\begin{equation}
\label{eq:Wigner:time}
W^{(e)}_{\rho,x}(t,\omega)=\int v_F\,
\mathcal{G}_{\rho,x}^{(e)}\left(t+\frac{\tau}{2},t-\frac{\tau}{2}\right)\,e^{i\omega\tau}\mathrm{d}\tau\,
\end{equation}
provides a very convenient real-valued representation for the single electron coherence. It describes the
nature of excitations (electron-like for $\omega>0$ and hole-like for $\omega<0$) as well as real
time aspects: both the electron distribution function $f_e(\omega)$ and the average electrical 
current $\langle i(x,t)\rangle$ are marginal distributions of the Wigner function, respectively obtained by averaging
over time or integrating over $\omega$~\cite{Ferraro:2013-1}. 
The Wigner function contains a Fermi sea contribution ensuring that
$W^{(e)}(t,\omega)\rightarrow 1$ when $\omega\rightarrow -\infty$ and the excess Wigner function 
$\Delta W^{(e)}_{\rho,x}(t,\omega)$
defined by subtracting the Fermi distribution function   
at the chemical potential at position $x$ thus represents the contribution of excitations.

\begin{figure*}
    \centering
\begingroup
  \makeatletter
  \providecommand\color[2][]{%
    \GenericError{(gnuplot) \space\space\space\@spaces}{%
      Package color not loaded in conjunction with
      terminal option `colourtext'%
    }{See the gnuplot documentation for explanation.%
    }{Either use 'blacktext' in gnuplot or load the package
      color.sty in LaTeX.}%
    \renewcommand\color[2][]{}%
  }%
  \providecommand\includegraphics[2][]{%
    \GenericError{(gnuplot) \space\space\space\@spaces}{%
      Package graphicx or graphics not loaded%
    }{See the gnuplot documentation for explanation.%
    }{The gnuplot epslatex terminal needs graphicx.sty or graphics.sty.}%
    \renewcommand\includegraphics[2][]{}%
  }%
  \providecommand\rotatebox[2]{#2}%
  \@ifundefined{ifGPcolor}{%
    \newif\ifGPcolor
    \GPcolorfalse
  }{}%
  \@ifundefined{ifGPblacktext}{%
    \newif\ifGPblacktext
    \GPblacktexttrue
  }{}%
  \let\gplgaddtomacro\g@addto@macro
  \gdef\gplbacktext{}%
  \gdef\gplfronttext{}%
  \makeatother
  \ifGPblacktext
    \def\colorrgb#1{}%
    \def\colorgray#1{}%
  \else
    \ifGPcolor
      \def\colorrgb#1{\color[rgb]{#1}}%
      \def\colorgray#1{\color[gray]{#1}}%
      \expandafter\def\csname LTw\endcsname{\color{white}}%
      \expandafter\def\csname LTb\endcsname{\color{black}}%
      \expandafter\def\csname LTa\endcsname{\color{black}}%
      \expandafter\def\csname LT0\endcsname{\color[rgb]{1,0,0}}%
      \expandafter\def\csname LT1\endcsname{\color[rgb]{0,1,0}}%
      \expandafter\def\csname LT2\endcsname{\color[rgb]{0,0,1}}%
      \expandafter\def\csname LT3\endcsname{\color[rgb]{1,0,1}}%
      \expandafter\def\csname LT4\endcsname{\color[rgb]{0,1,1}}%
      \expandafter\def\csname LT5\endcsname{\color[rgb]{1,1,0}}%
      \expandafter\def\csname LT6\endcsname{\color[rgb]{0,0,0}}%
      \expandafter\def\csname LT7\endcsname{\color[rgb]{1,0.3,0}}%
      \expandafter\def\csname LT8\endcsname{\color[rgb]{0.5,0.5,0.5}}%
    \else
      \def\colorrgb#1{\color{black}}%
      \def\colorgray#1{\color[gray]{#1}}%
      \expandafter\def\csname LTw\endcsname{\color{white}}%
      \expandafter\def\csname LTb\endcsname{\color{black}}%
      \expandafter\def\csname LTa\endcsname{\color{black}}%
      \expandafter\def\csname LT0\endcsname{\color{black}}%
      \expandafter\def\csname LT1\endcsname{\color{black}}%
      \expandafter\def\csname LT2\endcsname{\color{black}}%
      \expandafter\def\csname LT3\endcsname{\color{black}}%
      \expandafter\def\csname LT4\endcsname{\color{black}}%
      \expandafter\def\csname LT5\endcsname{\color{black}}%
      \expandafter\def\csname LT6\endcsname{\color{black}}%
      \expandafter\def\csname LT7\endcsname{\color{black}}%
      \expandafter\def\csname LT8\endcsname{\color{black}}%
    \fi
  \fi
  \setlength{\unitlength}{0.0500bp}%
  \begin{picture}(9070.00,5668.00)%
    \gplgaddtomacro\gplbacktext{%
      \csname LTb\endcsname%
      \put(1405,5431){\makebox(0,0){\strut{}(a) $\tau_s= 0$}}%
    }%
    \gplgaddtomacro\gplfronttext{%
      \csname LTb\endcsname%
      \put(765,3286){\makebox(0,0){\strut{}-2}}%
      \put(1085,3286){\makebox(0,0){\strut{} 0}}%
      \put(1405,3286){\makebox(0,0){\strut{} 2}}%
      \put(1725,3286){\makebox(0,0){\strut{} 4}}%
      \put(2045,3286){\makebox(0,0){\strut{} 6}}%
      \put(1405,2956){\makebox(0,0){\strut{}$(t-\bar\tau)/\tau_{\mathrm{e}}$}}%
      \put(433,3572){\makebox(0,0)[r]{\strut{}-5}}%
      \put(433,3878){\makebox(0,0)[r]{\strut{} 0}}%
      \put(433,4184){\makebox(0,0)[r]{\strut{} 5}}%
      \put(433,4488){\makebox(0,0)[r]{\strut{} 10}}%
      \put(433,4794){\makebox(0,0)[r]{\strut{} 15}}%
      \put(433,5100){\makebox(0,0)[r]{\strut{} 20}}%
      \put(-29,4336){\rotatebox{-270}{\makebox(0,0){\strut{}$\omega\tau_{\mathrm{e}}$}}}%
    }%
    \gplgaddtomacro\gplbacktext{%
      \csname LTb\endcsname%
      \put(4171,5431){\makebox(0,0){\strut{}(b) $\tau_s = 0.2\tau_{\mathrm{e}}$}}%
    }%
    \gplgaddtomacro\gplfronttext{%
      \csname LTb\endcsname%
      \put(3557,3286){\makebox(0,0){\strut{}-2}}%
      \put(3864,3286){\makebox(0,0){\strut{} 0}}%
      \put(4171,3286){\makebox(0,0){\strut{} 2}}%
      \put(4478,3286){\makebox(0,0){\strut{} 4}}%
      \put(4785,3286){\makebox(0,0){\strut{} 6}}%
      \put(4171,2956){\makebox(0,0){\strut{}$(t-\bar\tau)/\tau_{\mathrm{e}}$}}%
      \put(3232,3572){\makebox(0,0)[r]{\strut{}-5}}%
      \put(3232,3878){\makebox(0,0)[r]{\strut{} 0}}%
      \put(3232,4184){\makebox(0,0)[r]{\strut{} 5}}%
      \put(3232,4488){\makebox(0,0)[r]{\strut{} 10}}%
      \put(3232,4794){\makebox(0,0)[r]{\strut{} 15}}%
      \put(3232,5100){\makebox(0,0)[r]{\strut{} 20}}%
      \put(2770,4336){\rotatebox{-270}{\makebox(0,0){\strut{}$\omega\tau_{\mathrm{e}}$}}}%
    }%
    \gplgaddtomacro\gplbacktext{%
      \csname LTb\endcsname%
      \put(6983,5431){\makebox(0,0){\strut{}(c) $\tau_s = 0.4\tau_{\mathrm{e}}$}}%
    }%
    \gplgaddtomacro\gplfronttext{%
      \csname LTb\endcsname%
      \put(6369,3286){\makebox(0,0){\strut{}-2}}%
      \put(6676,3286){\makebox(0,0){\strut{} 0}}%
      \put(6983,3286){\makebox(0,0){\strut{} 2}}%
      \put(7290,3286){\makebox(0,0){\strut{} 4}}%
      \put(7597,3286){\makebox(0,0){\strut{} 6}}%
      \put(6983,2956){\makebox(0,0){\strut{}$(t-\bar\tau)/\tau_{\mathrm{e}}$}}%
      \put(6044,3572){\makebox(0,0)[r]{\strut{}-5}}%
      \put(6044,3878){\makebox(0,0)[r]{\strut{} 0}}%
      \put(6044,4184){\makebox(0,0)[r]{\strut{} 5}}%
      \put(6044,4488){\makebox(0,0)[r]{\strut{} 10}}%
      \put(6044,4794){\makebox(0,0)[r]{\strut{} 15}}%
      \put(6044,5100){\makebox(0,0)[r]{\strut{} 20}}%
      \put(5582,4336){\rotatebox{-270}{\makebox(0,0){\strut{}$\omega\tau_{\mathrm{e}}$}}}%
      \put(8793,567){\makebox(0,0)[l]{\strut{}-0.4}}%
      \put(8793,1020){\makebox(0,0)[l]{\strut{}-0.2}}%
      \put(8793,1473){\makebox(0,0)[l]{\strut{} 0}}%
      \put(8793,1927){\makebox(0,0)[l]{\strut{} 0.2}}%
      \put(8793,2380){\makebox(0,0)[l]{\strut{} 0.4}}%
      \put(8793,2834){\makebox(0,0)[l]{\strut{} 0.6}}%
      \put(8793,3287){\makebox(0,0)[l]{\strut{} 0.8}}%
      \put(8793,3740){\makebox(0,0)[l]{\strut{} 1}}%
      \put(8793,4194){\makebox(0,0)[l]{\strut{} 1.2}}%
      \put(8793,4647){\makebox(0,0)[l]{\strut{} 1.4}}%
      \put(8793,5101){\makebox(0,0)[l]{\strut{} 1.6}}%
    }%
    \gplgaddtomacro\gplbacktext{%
      \csname LTb\endcsname%
      \put(1360,2427){\makebox(0,0){\strut{}(d) $\tau_s= 0.6\tau_{\mathrm{e}}$}}%
    }%
    \gplgaddtomacro\gplfronttext{%
      \csname LTb\endcsname%
      \put(746,282){\makebox(0,0){\strut{}-2}}%
      \put(1053,282){\makebox(0,0){\strut{} 0}}%
      \put(1360,282){\makebox(0,0){\strut{} 2}}%
      \put(1667,282){\makebox(0,0){\strut{} 4}}%
      \put(1974,282){\makebox(0,0){\strut{} 6}}%
      \put(1360,-48){\makebox(0,0){\strut{}$(t-\bar\tau)/\tau_{\mathrm{e}}$}}%
      \put(421,568){\makebox(0,0)[r]{\strut{}-5}}%
      \put(421,874){\makebox(0,0)[r]{\strut{} 0}}%
      \put(421,1180){\makebox(0,0)[r]{\strut{} 5}}%
      \put(421,1484){\makebox(0,0)[r]{\strut{} 10}}%
      \put(421,1790){\makebox(0,0)[r]{\strut{} 15}}%
      \put(421,2096){\makebox(0,0)[r]{\strut{} 20}}%
      \put(-41,1332){\rotatebox{-270}{\makebox(0,0){\strut{}$\omega\tau_{\mathrm{e}}$}}}%
      \put(8793,567){\makebox(0,0)[l]{\strut{}-0.4}}%
      \put(8793,1020){\makebox(0,0)[l]{\strut{}-0.2}}%
      \put(8793,1473){\makebox(0,0)[l]{\strut{} 0}}%
      \put(8793,1927){\makebox(0,0)[l]{\strut{} 0.2}}%
      \put(8793,2380){\makebox(0,0)[l]{\strut{} 0.4}}%
      \put(8793,2834){\makebox(0,0)[l]{\strut{} 0.6}}%
      \put(8793,3287){\makebox(0,0)[l]{\strut{} 0.8}}%
      \put(8793,3740){\makebox(0,0)[l]{\strut{} 1}}%
      \put(8793,4194){\makebox(0,0)[l]{\strut{} 1.2}}%
      \put(8793,4647){\makebox(0,0)[l]{\strut{} 1.4}}%
      \put(8793,5101){\makebox(0,0)[l]{\strut{} 1.6}}%
    }%
    \gplgaddtomacro\gplbacktext{%
      \csname LTb\endcsname%
      \put(4171,2427){\makebox(0,0){\strut{}(e) $\tau_s = 0.8\tau_{\mathrm{e}}$}}%
    }%
    \gplgaddtomacro\gplfronttext{%
      \csname LTb\endcsname%
      \put(3557,282){\makebox(0,0){\strut{}-2}}%
      \put(3864,282){\makebox(0,0){\strut{} 0}}%
      \put(4171,282){\makebox(0,0){\strut{} 2}}%
      \put(4478,282){\makebox(0,0){\strut{} 4}}%
      \put(4785,282){\makebox(0,0){\strut{} 6}}%
      \put(4171,-48){\makebox(0,0){\strut{}$(t-\bar\tau)/\tau_{\mathrm{e}}$}}%
      \put(3232,568){\makebox(0,0)[r]{\strut{}-5}}%
      \put(3232,874){\makebox(0,0)[r]{\strut{} 0}}%
      \put(3232,1180){\makebox(0,0)[r]{\strut{} 5}}%
      \put(3232,1484){\makebox(0,0)[r]{\strut{} 10}}%
      \put(3232,1790){\makebox(0,0)[r]{\strut{} 15}}%
      \put(3232,2096){\makebox(0,0)[r]{\strut{} 20}}%
      \put(2770,1332){\rotatebox{-270}{\makebox(0,0){\strut{}$\omega\tau_{\mathrm{e}}$}}}%
      \put(8793,567){\makebox(0,0)[l]{\strut{}-0.4}}%
      \put(8793,1020){\makebox(0,0)[l]{\strut{}-0.2}}%
      \put(8793,1473){\makebox(0,0)[l]{\strut{} 0}}%
      \put(8793,1927){\makebox(0,0)[l]{\strut{} 0.2}}%
      \put(8793,2380){\makebox(0,0)[l]{\strut{} 0.4}}%
      \put(8793,2834){\makebox(0,0)[l]{\strut{} 0.6}}%
      \put(8793,3287){\makebox(0,0)[l]{\strut{} 0.8}}%
      \put(8793,3740){\makebox(0,0)[l]{\strut{} 1}}%
      \put(8793,4194){\makebox(0,0)[l]{\strut{} 1.2}}%
      \put(8793,4647){\makebox(0,0)[l]{\strut{} 1.4}}%
      \put(8793,5101){\makebox(0,0)[l]{\strut{} 1.6}}%
    }%
    \gplgaddtomacro\gplbacktext{%
      \csname LTb\endcsname%
      \put(6983,2427){\makebox(0,0){\strut{}(f) $\tau_s= \tau_{\mathrm{e}}$}}%
    }%
    \gplgaddtomacro\gplfronttext{%
      \csname LTb\endcsname%
      \put(6369,282){\makebox(0,0){\strut{}-2}}%
      \put(6676,282){\makebox(0,0){\strut{} 0}}%
      \put(6983,282){\makebox(0,0){\strut{} 2}}%
      \put(7290,282){\makebox(0,0){\strut{} 4}}%
      \put(7597,282){\makebox(0,0){\strut{} 6}}%
      \put(6983,-48){\makebox(0,0){\strut{}$(t-\bar\tau)/\tau_{\mathrm{e}}$}}%
      \put(6044,568){\makebox(0,0)[r]{\strut{}-5}}%
      \put(6044,874){\makebox(0,0)[r]{\strut{} 0}}%
      \put(6044,1180){\makebox(0,0)[r]{\strut{} 5}}%
      \put(6044,1484){\makebox(0,0)[r]{\strut{} 10}}%
      \put(6044,1790){\makebox(0,0)[r]{\strut{} 15}}%
      \put(6044,2096){\makebox(0,0)[r]{\strut{} 20}}%
      \put(5582,1332){\rotatebox{-270}{\makebox(0,0){\strut{}$\omega\tau_{\mathrm{e}}$}}}%
      \put(8793,567){\makebox(0,0)[l]{\strut{}-0.4}}%
      \put(8793,1020){\makebox(0,0)[l]{\strut{}-0.2}}%
      \put(8793,1473){\makebox(0,0)[l]{\strut{} 0}}%
      \put(8793,1927){\makebox(0,0)[l]{\strut{} 0.2}}%
      \put(8793,2380){\makebox(0,0)[l]{\strut{} 0.4}}%
      \put(8793,2834){\makebox(0,0)[l]{\strut{} 0.6}}%
      \put(8793,3287){\makebox(0,0)[l]{\strut{} 0.8}}%
      \put(8793,3740){\makebox(0,0)[l]{\strut{} 1}}%
      \put(8793,4194){\makebox(0,0)[l]{\strut{} 1.2}}%
      \put(8793,4647){\makebox(0,0)[l]{\strut{} 1.4}}%
      \put(8793,5101){\makebox(0,0)[l]{\strut{} 1.6}}%
    }%
    \gplbacktext
    \put(0,0){\includegraphics[width=16cm]{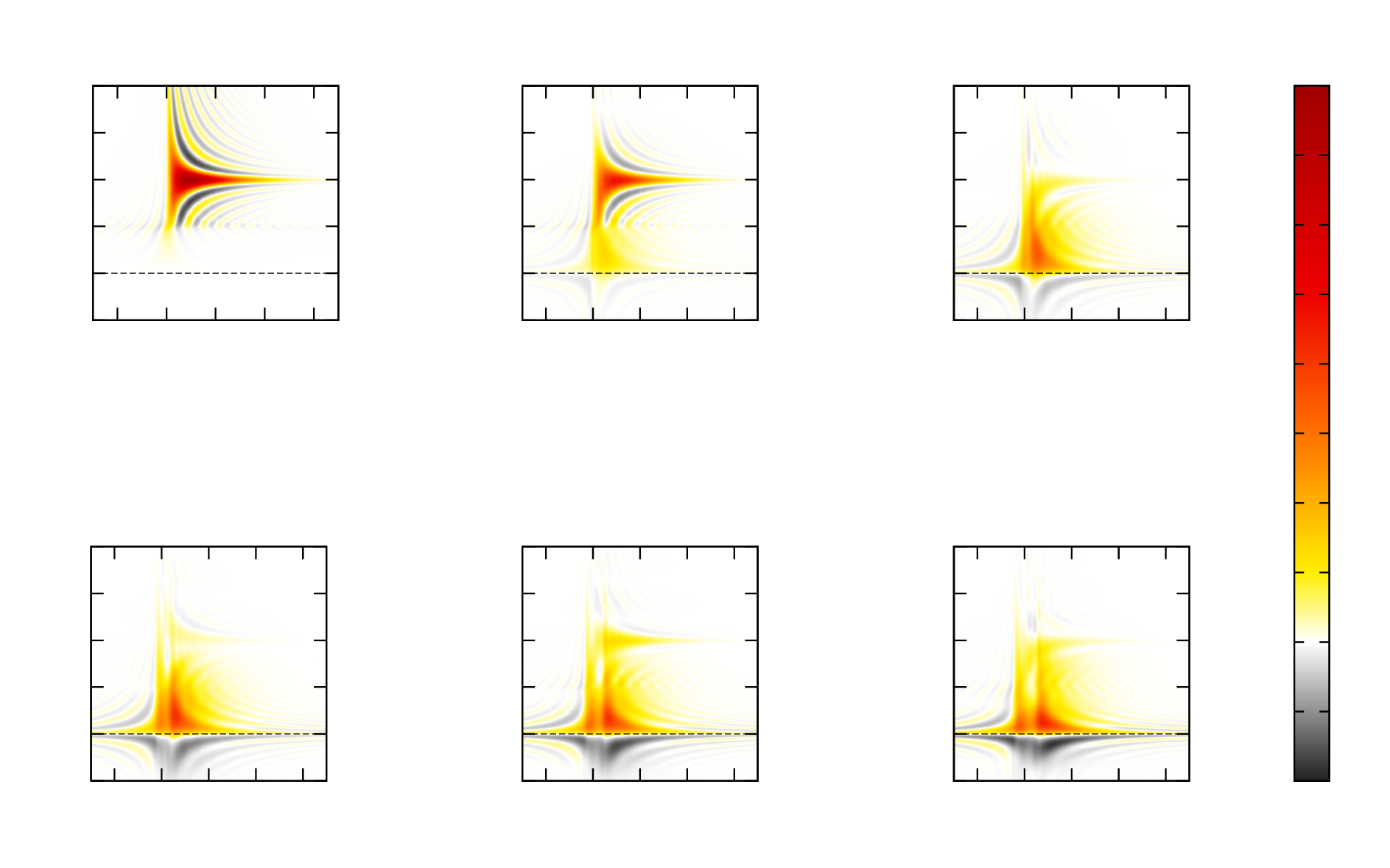}}%
    \gplfronttext
  \end{picture}%
\endgroup
 \caption{\label{fig:relaxation:Landau} (Color online)
Excess Wigner function $\Delta W^{(e)}_{\mathrm{out}}(t,\omega)$ at various propagation times for a Lorentzian wave packet in energy (Landau excitation)
of lifetime $\tau_{\mathrm{e}}=\gamma_{\mathrm{e}}^{-1}$ emitted 
at energy $\hbar \omega_{e}=10\hbar\tau_{\mathrm{e}}^{-1}$. The Fermi level
is indicated as an horizontal dashed  line at $\omega=0$. The various plots correspond to increasing propagating lengths
expressed in terms of times of flight: $\tau_s$ for the slow mode and $\tau_{c}=\tau_{s}/20$ corresponds to the
fast mode. The time shift $\bar{\tau}=(\tau_{c}+\tau_{s})/2$ compensates for the global drift of the excitations.
The initial excess Wigner function is depicted on panel (a).
Most of the electronic relaxation takes place at short times of flight (panels (a) to (c)) whereas the spin-charge separation
appears over longer times of flight (panels (d) to (f)).}
\end{figure*}

\begin{figure*}
\begingroup
  \makeatletter
  \providecommand\color[2][]{%
    \GenericError{(gnuplot) \space\space\space\@spaces}{%
      Package color not loaded in conjunction with
      terminal option `colourtext'%
    }{See the gnuplot documentation for explanation.%
    }{Either use 'blacktext' in gnuplot or load the package
      color.sty in LaTeX.}%
    \renewcommand\color[2][]{}%
  }%
  \providecommand\includegraphics[2][]{%
    \GenericError{(gnuplot) \space\space\space\@spaces}{%
      Package graphicx or graphics not loaded%
    }{See the gnuplot documentation for explanation.%
    }{The gnuplot epslatex terminal needs graphicx.sty or graphics.sty.}%
    \renewcommand\includegraphics[2][]{}%
  }%
  \providecommand\rotatebox[2]{#2}%
  \@ifundefined{ifGPcolor}{%
    \newif\ifGPcolor
    \GPcolorfalse
  }{}%
  \@ifundefined{ifGPblacktext}{%
    \newif\ifGPblacktext
    \GPblacktexttrue
  }{}%
  \let\gplgaddtomacro\g@addto@macro
  \gdef\gplbacktext{}%
  \gdef\gplfronttext{}%
  \makeatother
  \ifGPblacktext
    \def\colorrgb#1{}%
    \def\colorgray#1{}%
  \else
    \ifGPcolor
      \def\colorrgb#1{\color[rgb]{#1}}%
      \def\colorgray#1{\color[gray]{#1}}%
      \expandafter\def\csname LTw\endcsname{\color{white}}%
      \expandafter\def\csname LTb\endcsname{\color{black}}%
      \expandafter\def\csname LTa\endcsname{\color{black}}%
      \expandafter\def\csname LT0\endcsname{\color[rgb]{1,0,0}}%
      \expandafter\def\csname LT1\endcsname{\color[rgb]{0,1,0}}%
      \expandafter\def\csname LT2\endcsname{\color[rgb]{0,0,1}}%
      \expandafter\def\csname LT3\endcsname{\color[rgb]{1,0,1}}%
      \expandafter\def\csname LT4\endcsname{\color[rgb]{0,1,1}}%
      \expandafter\def\csname LT5\endcsname{\color[rgb]{1,1,0}}%
      \expandafter\def\csname LT6\endcsname{\color[rgb]{0,0,0}}%
      \expandafter\def\csname LT7\endcsname{\color[rgb]{1,0.3,0}}%
      \expandafter\def\csname LT8\endcsname{\color[rgb]{0.5,0.5,0.5}}%
    \else
      \def\colorrgb#1{\color{black}}%
      \def\colorgray#1{\color[gray]{#1}}%
      \expandafter\def\csname LTw\endcsname{\color{white}}%
      \expandafter\def\csname LTb\endcsname{\color{black}}%
      \expandafter\def\csname LTa\endcsname{\color{black}}%
      \expandafter\def\csname LT0\endcsname{\color{black}}%
      \expandafter\def\csname LT1\endcsname{\color{black}}%
      \expandafter\def\csname LT2\endcsname{\color{black}}%
      \expandafter\def\csname LT3\endcsname{\color{black}}%
      \expandafter\def\csname LT4\endcsname{\color{black}}%
      \expandafter\def\csname LT5\endcsname{\color{black}}%
      \expandafter\def\csname LT6\endcsname{\color{black}}%
      \expandafter\def\csname LT7\endcsname{\color{black}}%
      \expandafter\def\csname LT8\endcsname{\color{black}}%
    \fi
  \fi
  \setlength{\unitlength}{0.0500bp}%
  \begin{picture}(9070.00,2834.00)%
    \gplgaddtomacro\gplbacktext{%
      \csname LTb\endcsname%
      \put(1405,2881){\makebox(0,0){\strut{}(a) $\tau_s = 0$}}%
    }%
    \gplgaddtomacro\gplfronttext{%
      \csname LTb\endcsname%
      \put(872,282){\makebox(0,0){\strut{}-20}}%
      \put(1405,282){\makebox(0,0){\strut{} 0}}%
      \put(1938,282){\makebox(0,0){\strut{} 20}}%
      \put(1405,-48){\makebox(0,0){\strut{}$(t-\bar{\tau})/\tau_0$}}%
      \put(433,568){\makebox(0,0)[r]{\strut{}-1}}%
      \put(433,1064){\makebox(0,0)[r]{\strut{} 0}}%
      \put(433,1559){\makebox(0,0)[r]{\strut{} 1}}%
      \put(433,2054){\makebox(0,0)[r]{\strut{} 2}}%
      \put(433,2550){\makebox(0,0)[r]{\strut{} 3}}%
      \put(103,1559){\rotatebox{-270}{\makebox(0,0){\strut{}$\omega\tau_0$}}}%
    }%
    \gplgaddtomacro\gplbacktext{%
      \csname LTb\endcsname%
      \put(4171,2881){\makebox(0,0){\strut{}(b) $\tau_s = 10\tau_0$}}%
    }%
    \gplgaddtomacro\gplfronttext{%
      \csname LTb\endcsname%
      \put(3660,282){\makebox(0,0){\strut{}-20}}%
      \put(4171,282){\makebox(0,0){\strut{} 0}}%
      \put(4682,282){\makebox(0,0){\strut{} 20}}%
      \put(4171,-48){\makebox(0,0){\strut{}$(t-\bar{\tau})/\tau_0$}}%
      \put(3232,568){\makebox(0,0)[r]{\strut{}-1}}%
      \put(3232,1064){\makebox(0,0)[r]{\strut{} 0}}%
      \put(3232,1559){\makebox(0,0)[r]{\strut{} 1}}%
      \put(3232,2054){\makebox(0,0)[r]{\strut{} 2}}%
      \put(3232,2550){\makebox(0,0)[r]{\strut{} 3}}%
      \put(2902,1559){\rotatebox{-270}{\makebox(0,0){\strut{}$\omega\tau_0$}}}%
    }%
    \gplgaddtomacro\gplbacktext{%
      \csname LTb\endcsname%
      \put(6983,2881){\makebox(0,0){\strut{}(c) $\tau_s = 20\tau_0$}}%
    }%
    \gplgaddtomacro\gplfronttext{%
      \csname LTb\endcsname%
      \put(6472,282){\makebox(0,0){\strut{}-20}}%
      \put(6983,282){\makebox(0,0){\strut{} 0}}%
      \put(7494,282){\makebox(0,0){\strut{} 20}}%
      \put(6983,-48){\makebox(0,0){\strut{}$(t-\bar{\tau})/\tau_0$}}%
      \put(6044,568){\makebox(0,0)[r]{\strut{}-1}}%
      \put(6044,1064){\makebox(0,0)[r]{\strut{} 0}}%
      \put(6044,1559){\makebox(0,0)[r]{\strut{} 1}}%
      \put(6044,2054){\makebox(0,0)[r]{\strut{} 2}}%
      \put(6044,2550){\makebox(0,0)[r]{\strut{} 3}}%
      \put(5714,1559){\rotatebox{-270}{\makebox(0,0){\strut{}$\omega\tau_0$}}}%
      \put(8793,897){\makebox(0,0)[l]{\strut{} 0}}%
      \put(8793,1558){\makebox(0,0)[l]{\strut{} 1}}%
      \put(8793,2219){\makebox(0,0)[l]{\strut{} 2}}%
    }%
    \gplbacktext
    \put(0,0){\includegraphics[width=16cm]{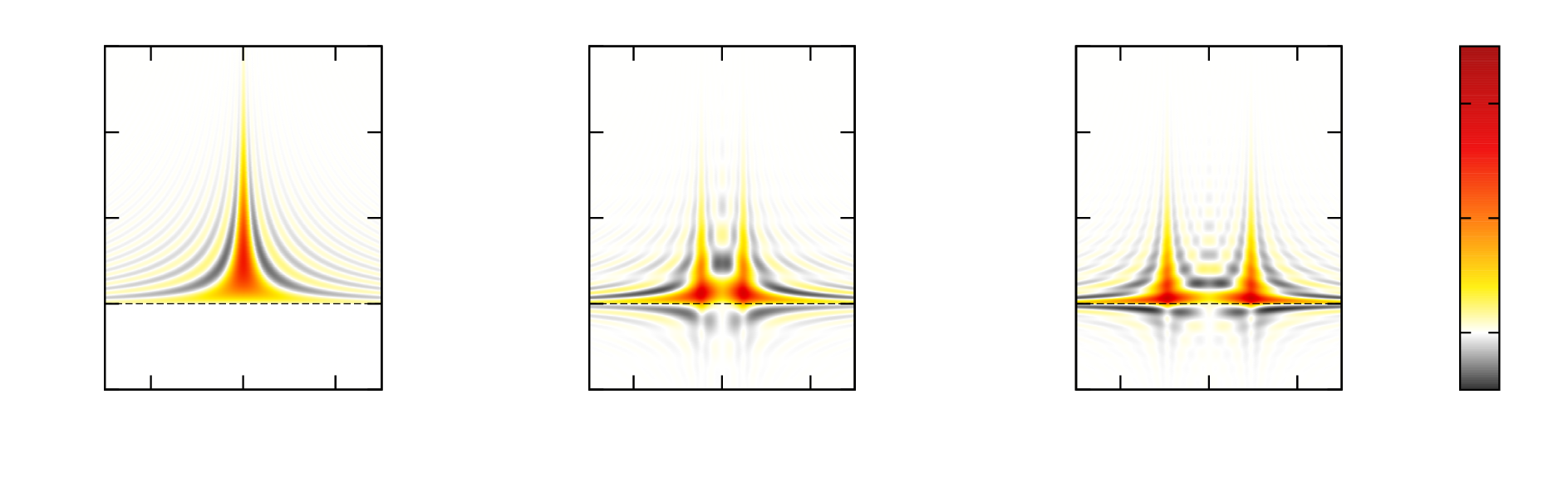}}%
    \gplfronttext
  \end{picture}%
\endgroup
\caption{\label{fig:relaxation:Leviton} (Color online) Excess Wigner function $\Delta W^{(e)}_{\mathrm{out}}(t,\omega)$ of a single electronic
Leviton of width $\tau_{0}$ for various propagation times: $\tau_{s}$ (resp. $\tau_c$) denotes the time of flight of the spin (resp. charged) mode
within the interaction region and $\tau_c=\tau_s/20$ and $\bar\tau=(\tau_s+\tau_c)/2$. The incoming Leviton splits into two half-Levitons which are
collective excitations.}
\end{figure*}

\begin{figure}
\begingroup
  \makeatletter
  \providecommand\color[2][]{%
    \GenericError{(gnuplot) \space\space\space\@spaces}{%
      Package color not loaded in conjunction with
      terminal option `colourtext'%
    }{See the gnuplot documentation for explanation.%
    }{Either use 'blacktext' in gnuplot or load the package
      color.sty in LaTeX.}%
    \renewcommand\color[2][]{}%
  }%
  \providecommand\includegraphics[2][]{%
    \GenericError{(gnuplot) \space\space\space\@spaces}{%
      Package graphicx or graphics not loaded%
    }{See the gnuplot documentation for explanation.%
    }{The gnuplot epslatex terminal needs graphicx.sty or graphics.sty.}%
    \renewcommand\includegraphics[2][]{}%
  }%
  \providecommand\rotatebox[2]{#2}%
  \@ifundefined{ifGPcolor}{%
    \newif\ifGPcolor
    \GPcolorfalse
  }{}%
  \@ifundefined{ifGPblacktext}{%
    \newif\ifGPblacktext
    \GPblacktexttrue
  }{}%
  \let\gplgaddtomacro\g@addto@macro
  \gdef\gplbacktext{}%
  \gdef\gplfronttext{}%
  \makeatother
  \ifGPblacktext
    \def\colorrgb#1{}%
    \def\colorgray#1{}%
  \else
    \ifGPcolor
      \def\colorrgb#1{\color[rgb]{#1}}%
      \def\colorgray#1{\color[gray]{#1}}%
      \expandafter\def\csname LTw\endcsname{\color{white}}%
      \expandafter\def\csname LTb\endcsname{\color{black}}%
      \expandafter\def\csname LTa\endcsname{\color{black}}%
      \expandafter\def\csname LT0\endcsname{\color[rgb]{1,0,0}}%
      \expandafter\def\csname LT1\endcsname{\color[rgb]{0,1,0}}%
      \expandafter\def\csname LT2\endcsname{\color[rgb]{0,0,1}}%
      \expandafter\def\csname LT3\endcsname{\color[rgb]{1,0,1}}%
      \expandafter\def\csname LT4\endcsname{\color[rgb]{0,1,1}}%
      \expandafter\def\csname LT5\endcsname{\color[rgb]{1,1,0}}%
      \expandafter\def\csname LT6\endcsname{\color[rgb]{0,0,0}}%
      \expandafter\def\csname LT7\endcsname{\color[rgb]{1,0.3,0}}%
      \expandafter\def\csname LT8\endcsname{\color[rgb]{0.5,0.5,0.5}}%
    \else
      \def\colorrgb#1{\color{black}}%
      \def\colorgray#1{\color[gray]{#1}}%
      \expandafter\def\csname LTw\endcsname{\color{white}}%
      \expandafter\def\csname LTb\endcsname{\color{black}}%
      \expandafter\def\csname LTa\endcsname{\color{black}}%
      \expandafter\def\csname LT0\endcsname{\color{black}}%
      \expandafter\def\csname LT1\endcsname{\color{black}}%
      \expandafter\def\csname LT2\endcsname{\color{black}}%
      \expandafter\def\csname LT3\endcsname{\color{black}}%
      \expandafter\def\csname LT4\endcsname{\color{black}}%
      \expandafter\def\csname LT5\endcsname{\color{black}}%
      \expandafter\def\csname LT6\endcsname{\color{black}}%
      \expandafter\def\csname LT7\endcsname{\color{black}}%
      \expandafter\def\csname LT8\endcsname{\color{black}}%
    \fi
  \fi
  \setlength{\unitlength}{0.0500bp}%
  \begin{picture}(4534.00,3968.00)%
    \gplgaddtomacro\gplbacktext{%
      \csname LTb\endcsname%
      \put(321,2097){\makebox(0,0)[r]{\strut{} 0}}%
      \put(321,2686){\makebox(0,0)[r]{\strut{} 0.4}}%
      \put(321,3275){\makebox(0,0)[r]{\strut{} 0.8}}%
      \put(824,3790){\makebox(0,0){\strut{} 0}}%
      \put(1288,3790){\makebox(0,0){\strut{} 1}}%
      \put(1751,3790){\makebox(0,0){\strut{} 2}}%
      \put(2215,3790){\makebox(0,0){\strut{} 3}}%
      \put(-119,2796){\rotatebox{-270}{\makebox(0,0){\strut{}$\langle i(t)\rangle$}}}%
      \put(1450,4009){\makebox(0,0){\strut{}$(t-\bar\tau)/\tau_e$}}%
      \put(592,3349){\makebox(0,0)[l]{\strut{}(a)}}%
    }%
    \gplgaddtomacro\gplfronttext{%
      \csname LTb\endcsname%
      \put(1988,3313){\makebox(0,0)[r]{\strut{}$\tau_s/\tau_{\mathrm{e}}=0$}}%
      \csname LTb\endcsname%
      \put(1988,3093){\makebox(0,0)[r]{\strut{}$0.2$}}%
      \csname LTb\endcsname%
      \put(1988,2873){\makebox(0,0)[r]{\strut{}$0.4$}}%
    }%
    \gplgaddtomacro\gplbacktext{%
      \csname LTb\endcsname%
      \put(2819,3790){\makebox(0,0){\strut{} 0}}%
      \put(3283,3790){\makebox(0,0){\strut{} 1}}%
      \put(3746,3790){\makebox(0,0){\strut{} 2}}%
      \put(4210,3790){\makebox(0,0){\strut{} 3}}%
      \put(3445,4009){\makebox(0,0){\strut{}$(t-\bar\tau)/\tau_e$}}%
      \put(2587,3349){\makebox(0,0)[l]{\strut{}(b)}}%
    }%
    \gplgaddtomacro\gplfronttext{%
      \csname LTb\endcsname%
      \put(3983,3313){\makebox(0,0)[r]{\strut{}$\tau_s/\tau_{\mathrm{e}}=0.6$}}%
      \csname LTb\endcsname%
      \put(3983,3093){\makebox(0,0)[r]{\strut{}$0.8$}}%
      \csname LTb\endcsname%
      \put(3983,2873){\makebox(0,0)[r]{\strut{}$1$}}%
    }%
    \gplgaddtomacro\gplbacktext{%
      \csname LTb\endcsname%
      \put(321,818){\makebox(0,0)[r]{\strut{} 0}}%
      \put(321,1380){\makebox(0,0)[r]{\strut{} 0.4}}%
      \put(321,1943){\makebox(0,0)[r]{\strut{} 0.8}}%
      \put(643,176){\makebox(0,0){\strut{}-5}}%
      \put(1118,176){\makebox(0,0){\strut{} 0}}%
      \put(1592,176){\makebox(0,0){\strut{} 5}}%
      \put(2067,176){\makebox(0,0){\strut{} 10}}%
      \put(-119,1169){\rotatebox{-270}{\makebox(0,0){\strut{}$\delta f_e(\omega)$}}}%
      \put(1450,0){\makebox(0,0){\strut{}$\omega\tau_e$}}%
      \put(643,607){\makebox(0,0)[l]{\strut{}(c)}}%
    }%
    \gplgaddtomacro\gplfronttext{%
      \csname LTb\endcsname%
      \put(1513,1692){\makebox(0,0)[r]{\strut{}$\tau_s/\tau_{\mathrm{e}}=0$}}%
      \csname LTb\endcsname%
      \put(1513,1472){\makebox(0,0)[r]{\strut{}$0.2$}}%
      \csname LTb\endcsname%
      \put(1513,1252){\makebox(0,0)[r]{\strut{}$0.4$}}%
    }%
    \gplgaddtomacro\gplbacktext{%
      \csname LTb\endcsname%
      \put(2638,176){\makebox(0,0){\strut{}-5}}%
      \put(3113,176){\makebox(0,0){\strut{} 0}}%
      \put(3587,176){\makebox(0,0){\strut{} 5}}%
      \put(4062,176){\makebox(0,0){\strut{} 10}}%
      \put(3445,0){\makebox(0,0){\strut{}$\omega\tau_e$}}%
      \put(2638,607){\makebox(0,0)[l]{\strut{}(d)}}%
    }%
    \gplgaddtomacro\gplfronttext{%
      \csname LTb\endcsname%
      \put(3508,1692){\makebox(0,0)[r]{\strut{}$\tau_s/\tau_{\mathrm{e}}=0.6$}}%
      \csname LTb\endcsname%
      \put(3508,1472){\makebox(0,0)[r]{\strut{}$0.8$}}%
      \csname LTb\endcsname%
      \put(3508,1252){\makebox(0,0)[r]{\strut{}$1$}}%
    }%
    \gplbacktext
    \put(0,0){\includegraphics[width=8cm]{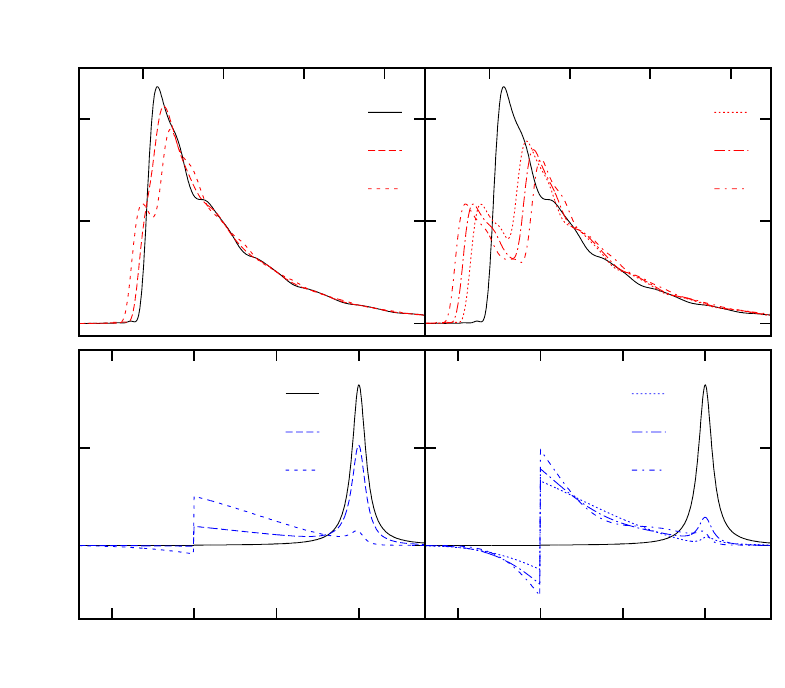}}%
    \gplfronttext
  \end{picture}%
\endgroup

\caption{\label{fig:marginals} (Color online) We present the
average current $\langle i(t)\rangle $ 
in units of $-e/\tau_{\mathrm{e}}$ (top row) and the
excess electron distribution function $\delta f_e(\omega)$ (bottom row)
at the same propagation distances as 
on Fig.~\ref{fig:relaxation:Landau} 
for an initial Lorentzian wave packet in energy (Landau excitation)
of lifetime $\tau_{\mathrm{e}}=\gamma_{\mathrm{e}}^{-1}$ and
energy $\hbar \omega_{e}=10\hbar\tau_{\mathrm{e}}^{-1}$.
The initial $\langle i(t)\rangle $  and $\delta f_e(\omega)$ appear as filled black curves on
all panels. For short propagation times (panels (a) and (c)), the current peaks are 
not very well separated and no hole excitations
are created but the quasi-particle peak in the electron distribution function strongly decays.
For longer propagation times (panels (b) and (d)), hole excitations 
appear as the current pulse fractionalizes
in two well separated peaks whereas the quasi-particle peak remains small.}
\end{figure}

Panel (a) of Fig.~\ref{fig:relaxation:Landau} presents the excess Wigner function of the Landau
single electron excitation of duration $\tau_{\mathrm{e}}=\gamma^{-1}_{\mathrm{e}}$
injected at an energy $\hbar\omega_{\mathrm{e}}$ above
the Fermi level. Since
$\omega_{\mathrm{e}}/\gamma_{\mathrm{e}}=10$, it is well separated from the Fermi sea. 
The corresponding Lorentzian peak of the excess electron distribution function $\delta f_e(\omega)$ can be
seen on panel (c) of Fig.~\ref{fig:marginals} while the corresponding current pulse is shown on panel (a).
By contrast, the excess Wigner function of a Levitov excitation depicted
on Panel (a) of Fig.~\ref{fig:relaxation:Leviton} is not well separated from the Fermi sea. Both
of them are purely electronic: they vanish in the half plane $\omega<0$ corresponding to hole excitations.

\medskip

Let us now send such a coherent single electron excitation through an interaction region of finite
length as shown on Fig.~\ref{fig:input-output}. 
Within this region, electrons experience screened Coulomb interactions with 
neighboring edge channels, other mesoscopic conductors  and gates. 
In many experimentally relevant cases, such effective interactions can be described 
in terms of edge-magnetoplasmon scattering~\cite{Safi:1995-1,Safi:1999-1,Degio:2010-1} which are
charge density waves along the edge channel. 
Assuming that all conductors involved are in the linear response regime when a single
electron excitation passes through, this scattering is elastic. 
In this simple picture depicted on Fig.~\ref{fig:input-output}, 
an incoming sinusoidal charge density wave at pulsation $\omega$ generates excitations at the same frequency in the
environmental modes through the capacitive coupling between the edge channel and its electromagnetic environment.
Under this hypothesis, the evolution
of the many-body state $|\varphi_{\mathrm{e}}, F\rangle $ can be completely
described in terms of the edge magnetoplasmon transmission amplitude $t(\omega)$~\cite{Degio:2009-1}. 
This also suggests that interactions are most conveniently handled in the frequency domain. 

Our key result is an analytical expression for the outcoming single 
electron coherence in the frequency domain 
$\widetilde{\mathcal{G}}^{(e)}_{\mathrm{out}}(\omega+\Omega/2,\omega-\Omega/2)$
where $\omega$ and $\Omega$ are respectively conjugated to $t-t'$ and $(t+t')/2$ in 
$\mathcal{G}_{\mathrm{out}}^{(e)}(t,t')$. The outcoming coherence is obtained 
in terms of the excess coherence 
 of the incoming wave packet
\begin{align}
\widetilde{\mathcal{G}}^{(e)}_{\mathrm{out}}\left(\omega+\frac{\Omega}{2},\omega-\frac{\Omega}{2}\right)
& = \nonumber \\ 
\int_{-\infty}^{+\infty} 
K\left(\omega,\omega';\Omega\right)\,
& \widetilde{\varphi}_{\mathrm{e}}\Bigl( \omega'+\frac{\Omega}{2}\Bigr)
\widetilde{\varphi}_{\mathrm{e}}^*\Bigl( \omega'-\frac{\Omega}{2}\Bigr) \mathrm{d}\omega'\,
\label{eq:out-coherence:frequency}
\end{align}
in which the propagator $K\left(\omega,\omega';\Omega\right)$ encodes 
the effect of interactions on single electron excitations and 
$\widetilde{\varphi}_{\mathrm{e}}(\omega)$
denotes the Fourier transform of $\varphi_{\mathrm{e}}(-v_Ft)$. Contrary to the
expressions  obtained for energy resolved single electron excitations~\cite{Degio:2009-1}, 
Eq.~\eqref{eq:out-coherence:frequency} and the analytical expressions for $K\left(\omega,\omega';\Omega\right)$ given
in appendix \ref{appendix:propagators}
describe both the time and energy dependence of electronic relaxation.
They provide the exact solution to the relaxation and decoherence problem for any incoming single electronic wave packet and
for generic effective screened Coulomb interactions described within the framework of elastic edge magnetoplasmon scattering.  In this Letter,
we focus on the case of the $\nu=2$ edge channel system with short range interactions, at strong coupling~\cite{Levkivskyi:2008-1} 
and zero temperature (see appendix \ref{appendix:interactions}).

\medskip

Numerical evaluations of the outcoming Wigner function for 
an incoming Lorentzian wavepacket in the energy domain centered around 
$\hbar\omega_{\mathrm{e}}$ and of width $\gamma_{\mathrm{e}}=\tau_{\mathrm{e}}^{-1}$ 
are depicted on Fig.~\ref{fig:relaxation:Landau} for various propagation distances or, equivalently, times
of flight. 
These results shed light on the decoherence scenario of the Landau quasi-particle: it clearly
involves a time scale separation. In the limit of an energy resolved wave packet	 
$\gamma_{\mathrm{e}}\ll \omega_{\mathrm{e}}$, the single electron coherence around
$\omega_{\mathrm{e}}$ relaxes close to the Fermi level
after a propagation distance proportional to $\omega_{\mathrm{e}}^{-1}$.
Then, after a propagation distance proportional to the 
wave packet duration $\gamma_{\mathrm{e}}^{-1}\gg \omega_{\mathrm{e}}^{-1}$, 
the single electron coherence splits into two parts
progressing at the velocities of the two edge magnetoplasmon eigenmodes, thus giving birth 
to collective excitations close to the Fermi sea. 
The first phenomenon is indeed associated with energy
relaxation probed by Le~Sueur {\it et al}~\cite{LeSueur:2010-1} 
whereas the second one corresponds to
the expected charged/neutral mode separation probed by Bocquillon 
{\it et al}~\cite{Bocquillon:2012-2}. This decoherence scenario should be compared to the one for the
Levitov quasiparticle shown on Fig.~\ref{fig:relaxation:Leviton}: the
Leviton splits into two half-Levitons which are Lorentzian current pulses carrying a charge $-e/2$. 
The only time scale appearing corresponds to the time needed to fractionalize a 
Leviton~\cite{Berg:2009-1,Neder:2012-1,Grenier:2013-1}.

\medskip

The difference between these two situations arises from the nature of the incoming many-body states:
for the Leviton, it is a coherent state of the edge magnetoplasmon modes or, equivalently, a quasi-classical 
charge density wave~\cite{Grenier:2013-1}. Since the interaction region acts as a frequency dependent beam
splitter for the edge magnetoplasmons, the Leviton many-body state does not entangle with environmental degrees
of freedom. In other words, it is a pointer state~\cite{Zurek:1993-1} which does not experience decoherence. The Wigner function 
changes shown on Fig.~\ref{fig:relaxation:Leviton} come from electron/hole pair generation 
in this pure many-body state.

On the other hand, the many-body state corresponding to the Landau quasi-particle is a coherent superposition
of such pointer states. It is thus subject to decoherence induced by entanglement with the second edge channel.
Depending on the copropagation distance, the outcoming many-body state of the edge channel under consideration is a partly 
decohered mixture of coherent states, each of them corresponding 
to a localized electronic excitation dressed by a cloud of electron-hole pairs (see 
appendix \ref{appendix:propagators}).
This extrinsic decoherence manifests itself through the rapid decay of the 
Wigner function at the initial injection energy $\hbar\omega_{\mathrm{e}}$. 

Therefore, the rapid electronic relaxation shown on Fig.~\ref{fig:relaxation:Landau} should be viewed as 
the electronic analogue of the decay of interference fringes expected for the Wigner function
of the superposition of two coherent states of an electromagnetic mode
recently observed in cavity QED experiments~\cite{Deleglise:2008-1}. Here, it 
arises from the decoherence of a mesoscopic superposition of quasi-classical charge density waves. 
This process takes place over a shorter time than the evolution of each of these quasi-classical states 
which corresponds to the spin-charge separation of the current pulses.
Once this decoherence process has taken place, the outcoming many-body state is
an incoherent mixture of fractionalized localized electronic excitations. We
confirm this scenario by computing both the current pulse and
the electron distribution function corresponding to each of the Wigner functions
of Fig.~\ref{fig:relaxation:Landau}: the decay of the quasi-particle peak takes 
place at short times (Fig.~\ref{fig:marginals}c).
During this time the current pulses are almost unseparated (Fig.~\ref{fig:marginals}a) and 
no hole excitations are created, thus confirming that it is a purely extrinsic 
decoherence effect. As the separation of the two half-charge current pulses becomes
more visible (Fig.~\ref{fig:marginals}b), hole excitations are created (Fig.~\ref{fig:marginals}d).

For typical duration of the wave packet generated by single electron sources~\cite{Bocquillon:2013-1} $\tau_{\mathrm{e}}\simeq
\unit{60}{\pico\second}$ and typical velocity of the slow mode~\cite{Bocquillon:2012-2} 
$v_{s}\simeq \unit{4.6\times 10^4}{\meter\usk\reciprocal\second}$, energy relaxation
takes place over a propagation distance smaller than $v_{s}\tau_{\mathrm{e}}\simeq \unit{2.8}{\micro\meter}$.
Therefore, without any decoherence/relaxation preserving design~\cite{Huynh:2012-1,Altimiras:2010-2}, the single electron coherence
measured after a few $\micro\meter$ propagation along a second edge channel 
corresponds to collective excitations close to the Fermi sea such as 
the one depicted on panel (f) of Fig.~\ref{fig:relaxation:Landau}. This decay of single electron coherence is responsible for
the reduction of contrast in MZI interferometry experiments~\cite{Levkivskyi:2008-1,Neuenhahn:2009-1}. Moreover, 
as pointed out by Wahl {\it et al} \cite{Wahl:2013-1}, interactions also lead to a reduction of the Pauli dip in HOM experiments, as 
recently observed~\cite{Bocquillon:2013-1}.

Since the experimental signal in an Hong-Ou-Mandel interferometer
is directly related to the overlap of the  two incoming excess Wigner functions~\cite{Ferraro:2013-1}, HOM experiments
can efficiently probe the outcoming single electron coherence with sub-nanosecond time resolution. 
First of all,  the side peak structures in the HOM signal predicted by Wahl {\it et al}~\cite{Wahl:2013-1}
reflect the splitting of the outcoming collective excitation Wigner function with increasing propagation distance
as can be seen on panels (d) to (f) of Fig.~\ref{fig:relaxation:Landau}.
But more generally, using for example very narrow Lorentzian 
pulses~\footnote{Minimizing the 
partition noise~\cite{Dubois:2013-1} or controlling the harmonics of the drive ensures that undistorted Levitons arrive at the QPC.} 
and a d.c. bias, an HOM-like experiment would probe the precise time and frequency dependence
of the outcoming Wigner function. Anti-Levitons (Lorentzian pulses carrying charge $+e$) could be used to probe both the time and
frequency dependence of the outcoming Wigner function for $\omega<0$ and the corresponding HOM signal may exhibit contributions
from the electron/hole coherences generated by interactions~\cite{Jonckheere:2012-1}. 
Finally, the edge magnetoplasmon decoherence manifests itself
through a striking phenomenon: once it has taken over, because all contributions arising from 
the initial coherence
$\varphi_{\mathrm{e}}(t_+)\varphi_{\mathrm{e}}^*(t_{-})$ at different times $t_{+}\neq t_{-}$ 
are suppressed, the 
outcoming single electron coherence only depends on $|\varphi_{\mathrm{e}}(t)|^{2}$. 
Consequently, it only depends on the shape of the incoming current pulse and
not on its injection energy. This striking feature can be easily tested in an HOM experiment: the HOM curve should not change
when modifying the injection energy of the electron while keeping the duration of the wave packet constant as long as 
it is injected well above the Fermi level. The precise shape of this curve does however depend on the 
form of effective interactions~\cite{Bocquillon:2012-2}. This dependence as well as the effect of non-zero  
temperature will be explored in a subsequent publication.

\medskip

To conclude, we have computed the Wigner function of a coherent single electron excitation after propagating across an interaction
region of the $\nu=2$ edge channel system. We have unraveled fundamental difference between the decoherence scenario
of the Landau and Levitov quasi-particles that reflects the decoherence of the 
corresponding final many-body states. We suggest that HOM type experiments could probe the incoherent mixture 
of fractionalized localized electronic excitations resulting from the effect of Coulomb interactions 
on the Landau quasi-particle. In particular, we propose a simple test of this decoherence based on the HOM
experiment. These experimental tests would provide useful information on the decoherence mechanisms and
effective interactions at a much lower experimental cost than the generic tomography protocols based on HBT~\cite{Degio:2010-4}
or MZI~\cite{Haack:2011-1} interferometry. 

\medskip

\acknowledgements{We thank J.M.~Berroir, E. Bocquillon, V. Freulon
and B.~Plaçais from LPA for useful discussions as well as C. Bauerle from Institut Neel. 
We also thank 
T.~Jonckheere, T.~Martin, J.~Rech and C.~Wahl from CPT Marseille for  
discussions about their work and for very useful remarks on our manuscript. 
This work is supported by the ANR grant ''1shot'' (ANR-2010-BLANC-0412).}


\appendix
\section{Interactions and edge-magnetoplasmon scattering}
\label{appendix:interactions}

In this paper, we consider the $\nu=2$ edge channel system and review the
description of screened Coulomb interactions in terms of edge magnetoplasmon scattering. To begin with, let us, in full generality, consider
a quantum Hall edge channel capacitively coupled to a linear environment such as, for example, an external gate
of size $l$ connected to an impedance representing a dissipative circuit~\cite{Degio:2009-1} or a second
edge channel~\cite{Levkivskyi:2008-1,Degio:2010-1}. The interaction region is assumed
to be of finite length. Solving the equations of motion for the bosonic degree of freedom describing the edge channel 
(edge magnetoplasmon modes) and the environmental bosonic modes then leads to an elastic scattering matrix that describes the effect of the interaction region
on both the edge-magnetoplasmon and the environmental modes. The precise form and frequency dependence of this
matrix depends on the precise form of effective Coulomb interactions used in the model. The edge magnetoplasmon scattering matrix is indeed related to
the finite frequency admittances of the conductors~\cite{Safi:1995-1,Safi:1999-1,Degio:2009-1}, thus enabling an experimental
determination of such plasmon scattering amplitudes~\cite{Bocquillon:2012-2}.

\medskip

\begin{figure*}
\begin{center}
\includegraphics[width=0.6\linewidth]{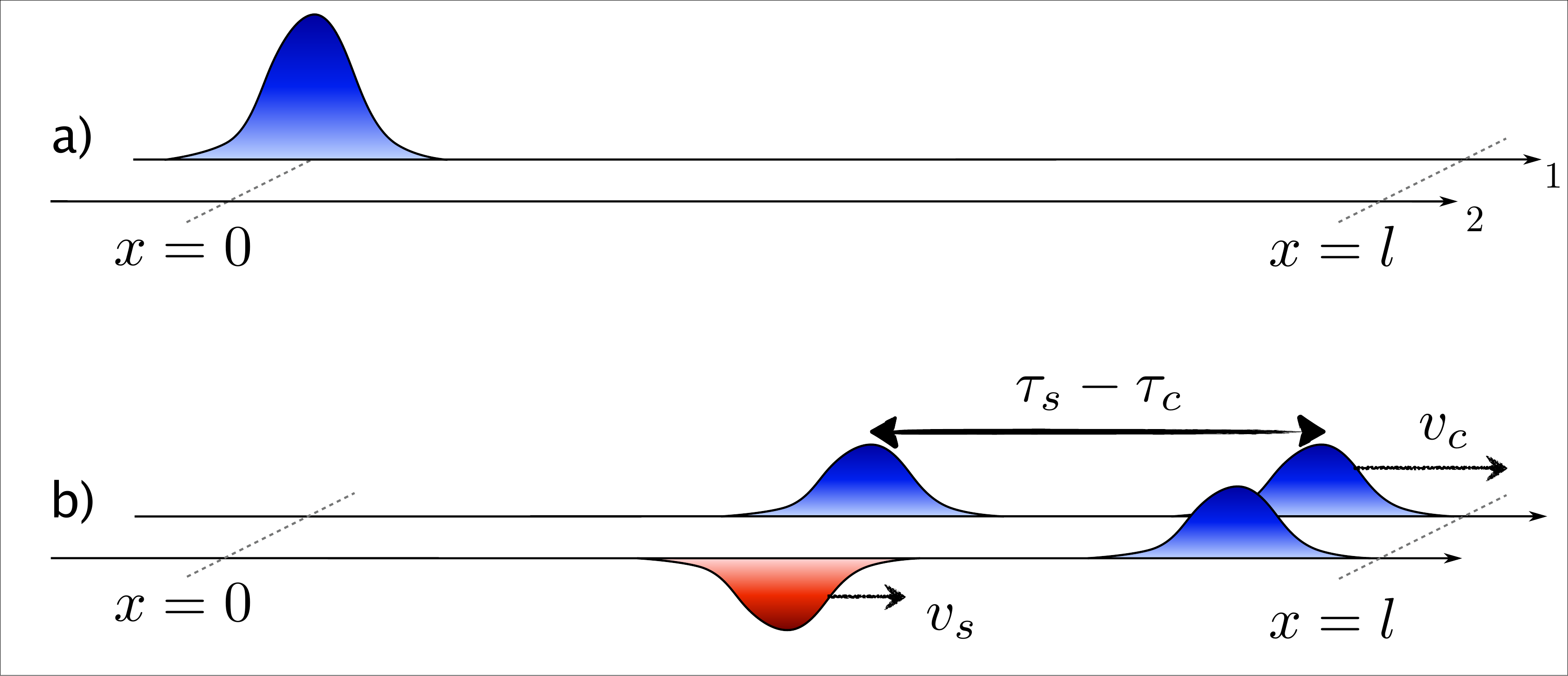}
 \caption{Propagation of a Lorentzian pulse in the $\nu=2$ edge channel system
 in the presence of short range interactions at strong coupling: (a) a Lorentzian pulse
 is injected on one of the two edge channels before the interaction region; (b) this excitation
 has been fractionalized leading to a fast moving charge excitation $c$ and a
 slow moving neutral excitation $s$.}
 \label{fig:neutral-charge}
 \end{center}
\end{figure*}

We now focus on two copropagating edge channels coupled over a distance $l$ by short range effective screened
Coulomb interactions. Over a copropagation distance $l$, such short range interactions
lead to the dispersionless propagation of the two edge magnetoplasmon modes which are called
the slow and the fast modes with respective velocities $v_s$ and $v_c$ ($v_c>v_s$). The interaction strength
is characterized by a single angular parameter $\theta$ which determines how these eigenmodes are
expressed as linear combinations of the edge magnetoplasmon modes of the two edge channels. Importantly,
the strong coupling regime corresponds to $\theta=\pi/2$~\cite{Levkivskyi:2008-1}. 
In this regime, the fast mode is symmetric and therefore
carries the total charge: this is why we denote it by the subscript $c$.
The slow mode is then neutral and corresponds to opposite charges on both edges. For this reason, the neutral
mode is often called dipolar or, assuming that both edge channels carry opposite
spins in the quantization direction given by the magnetic field, the spin mode. 
This leads to the following edge-magnetoplasmon scattering~\cite{Degio:2010-1}:
\begin{equation}
 S(\omega)=\begin{pmatrix}
            p_{s}\,e^{i\omega \tau_{s}} + 
            p_{c}\,e^{i\omega \tau_{c}}
            & q \left(e^{i\omega \tau_{c}}-
		      e^{i\omega \tau_{s}} \right)
            \\
            q \left(e^{i\omega \tau_{c}}-
		      e^{i\omega \tau_{s}} \right)
	    & p_{c}\,e^{i\omega \tau_{s}} + 
	      p_{s}\,e^{i\omega \tau_{c}}
           \end{pmatrix},
\end{equation}
where $\tau_s=l/v_{s}$  and  $\tau_{c}=l/v_{c}$ respectively denotes the time of flight
of the slow and fast modes and
\begin{align}
 p_{s/c}=\frac{1\pm\cos(\theta)}{2}, &\qquad q=\frac{\sin(\theta)}{2}\,.
\end{align}
Because the edge magnetoplasmon modes are dispersionless, the charge density wave generated
by applying a time dependent voltage drive to one of the two edge channels propagates in a very simple way:
a classical drive generates an edge magnetoplasmon coherent state which thus
propagates just as a classical electromagnetic wave through a frequency dependent beam splitter.
As a consequence, the outcoming quantum state is just a tensor product of two edge magnetoplasmon states
that correspond to the outcoming transmitted and reflected waves~\cite{Grenier:2013-1}. 
Fig.~\ref{fig:neutral-charge} depicts
the result of this process for an incoming Lorentzian pulse in the time or space domain 
such as, for example, the recently generated Levitov quasi-particle
or Leviton~\cite{Dubois:2013-2}:  due to the
absence of dispersion for the fast and slow eigenmodes~\cite{Grenier:2013-1}, 
this excitation splits into two Lorentzian pulses with half amplitude.

\medskip

Recently, finite frequency admittance measurements by 
Bocquillon {\it et al}~\cite{Bocquillon:2012-2} have shown that this picture is valid at low energies
and that the eigenmodes correspond to the strong coupling case $\theta= \pi/2$. 
Fractionalization has also been recently confirmed
by Heiblum {\it et al} through noise measurements but with a different mixing angle ($\theta\simeq \pi/3$)~\cite{Inoue:2013-1}. 
Nevertheless, we shall focus on the strong coupling case in this paper.

\section{Computing single electron coherence}
\label{appendix:propagators}

We shall now present the method used to compute the single electron excess coherence coming
out of the interaction region. The analytical expressions used for numerical evaluations are valid at zero temperature,
for arbitrary incoming single electron wave packet and for arbitrary effective interactions described by
edge magnetoplasmon scattering. We proceed by deriving numerical expressions for the excess single
electron coherence in the frequency domain. The excess Wigner function~\cite{Ferraro:2013-1} is then evaluated 
by performing a discrete Fourier transform (DFT).

\subsection{Computing the outcoming single electron coherence}

Let us consider an incoming many-body state of the form
\begin{equation}
|\varphi_{\mathrm{e}},F\rangle = v_F\int \varphi_{\mathrm{e}}(t)\psi^\dagger(t)\,|F\rangle\,\mathrm{d}t\,
\end{equation}
where $\varphi_{\mathrm{e}}(t)$ denotes the electronic wave packet evaluated at $x=-v_Ft$. The resulting incoming single electron coherence
is then obtained as a double integral involving a four point fermionic correlator:
\begin{widetext}
\begin{equation}
\label{eq:coherence:incoming:4pt}
\mathcal{G}^{(e)}_{\mathrm{out}}\left(t+\frac{\tau}{2},t-\frac{\tau}{2}\right) =  v_F^2
\int \varphi_{\mathrm{e}}(t_+)\varphi_{\mathrm{e}}(t_-)^*
\left\langle \psi(t_-)\psi^\dagger\left(t-\frac{\tau}{2}\right)\psi\left(t+\frac{\tau}{2}\right)
\psi^\dagger(t_+)\right\rangle_F\,\mathrm{d}t_+\mathrm{d}t_-\,
\end{equation}
\end{widetext}
where $\langle A\rangle_F$ denotes the quantum average of $A$ in the Fermi sea $|F\rangle$.
Using Wick's theorem, the incoming single
electron contribution is obtained as the sum of two contributions, one corresponding to 
the Fermi sea and one corresponding to the electronic wave packet:
\begin{widetext}
\begin{subequations}
\begin{eqnarray}
\label{eq:coherence:free:decomposition}
\mathcal{G}^{(e)}_{\mathrm{in}}\left(t+\frac{\tau}{2},t-\frac{\tau}{2}\right) & =  & 
\mathcal{G}^{(e)}_F(\tau) +
\mathcal{G}^{(e)}_{\mathrm{in},\mathrm{wp}}\left(t+\frac{\tau}{2},t-\frac{\tau}{2}\right)
\\
\mathcal{G}^{(e)}_{\mathrm{in},\mathrm{wp}}\left(t+\frac{\tau}{2},t-\frac{\tau}{2}\right) & =  & v_F^2 \int 
\varphi_{\mathrm{e}}(t_+)\varphi_{\mathrm{e}}(t_-)^*
\left\langle \psi\left(t+\frac{\tau}{2}\right)\,\psi^\dagger(t_+)\right\rangle_F\,
\left\langle \psi\left(t-\frac{\tau}{2}\right)\,\psi^\dagger(t_-)\right\rangle_F^*\,
\mathrm{d}t_+\mathrm{d}t_-\,
\label{eq:coherence:free:wp}
\end{eqnarray}
\end{subequations}
\end{widetext}
where $\mathcal{G}_F^{(e)}(\tau)=\langle \psi^\dagger(0)\psi(\tau)\rangle_F$ denotes the single electron coherence
of the Fermi sea.
Since we consider a purely electronic wave packet $\varphi_{\mathrm{e}}$ the two point fermionic correlators can safely be replaced
by $\delta$ functions, thus leading to the familiar expression for the excess single electron coherence 
$\Delta\mathcal{G}^{(e)}=\mathcal{G}^{(e)}
-\mathcal{G}^{(e)}_F$:
\begin{equation}
\Delta\mathcal{G}^{(e)}_{\mathrm{in}}
\left(t+\frac{\tau}{2},t-\frac{\tau}{2}\right)=\varphi_{\mathrm{e}}\left(t+\frac{\tau}{2}\right)\,
\varphi_{\mathrm{e}}\left(t-\frac{\tau}{2}\right)^*\,.
\end{equation}
In order to compute the outcoming single electron coherence within the bosonization
framework, we first trace out over the environmental degrees of freedom. The
outcoming single electron coherence is then equal to:
\begin{widetext}
\begin{equation}
\label{eq:Gout}
\mathcal{G}^{(e)}_{\mathrm{out}}\left(t+\frac{\tau}{2},t-\frac{\tau}{2}\right)
=v_F^2\int \varphi_{\mathrm{e}}(t_+)\varphi_{\mathrm{e}}(t_-)^*\,
\mathcal{D}_{\mathrm{ext}}(t_+-t_-)\,\left\langle g(t_-)\left|\psi(t_-)\psi^\dagger\left(t-\frac{\tau}{2}\right)\psi\left(t+\frac{\tau}{2}\right)
\psi^\dagger(t_+)\right|g(t_+)\right\rangle\,\mathrm{d}t_+\mathrm{d}t_-\,.
\end{equation}
\end{widetext}
where $\mathcal{D}_{\mathrm{ext}}(t_+-t_-)$ is the extrinsic decoherence coefficient that arises
from the imprints left in the external environment by localized fermionic excitations located at $t_+$ and
$t_-$. It is given in terms of the scattering probability $R(\omega)$ of an edge
magnetoplasmon into the environmental modes at the same frequency~\cite{Degio:2009-1}:
\begin{equation}
\mathcal{D}_{\mathrm{ext}}(\tau)=\exp{\left(\int_0^{+\infty}R(\omega)\,(e^{i\omega \tau}-1)\frac{\mathrm{d}\omega}{\omega}\right)}\,
\end{equation}
in which $R(\omega)=|S_{12}(\omega)|^2$.
The states $|g(t_\pm)\rangle$ in Eq.~\eqref{eq:Gout} are the edge magnetoplasmon coherent states describing the 
corresponding electron/hole pair clouds generated into the edge channels by screened Coulomb interactions:
\begin{equation}
|g(t_\pm)\rangle = \bigotimes_{\omega >0}\left|\frac{e^{i\omega t_\pm}}{\sqrt{\omega}}(1-t(\omega))\right\rangle \,.
\end{equation}
where $t(\omega)=S_{11}(\omega)$ denotes the edge magnetoplasmon transmission amplitude.
This finally 
leads to an expression for the outcoming single electron coherence that generalizes Eq.~\eqref{eq:coherence:incoming:4pt}
as:
\begin{widetext}
\begin{equation}
\label{eq:coherence:outcoming:4pt}
\mathcal{G}^{(e)}_{\mathrm{out}}\left(t+\frac{\tau}{2},t-\frac{\tau}{2}\right) = v_F^2
\int \varphi_{\mathrm{e}}(t_+)\varphi_{\mathrm{e}}(t_-)^*\left\langle \psi(t_-)\psi^\dagger\left(t-\frac{\tau}{2}\right)\psi\left(t+\frac{\tau}{2}\right)
\psi^\dagger(t_+)\right\rangle_F\, \mathcal{D}_{t_+,t_-}(t,\tau)\,\mathrm{d}t_+\mathrm{d}t_-\,
\end{equation}
\end{widetext}
where the full decoherence coefficient $ \mathcal{D}_{t_+,t_-}(t,\tau)$ takes into account both the effect
of extrinsic decoherence and of the overlap of the accompanying clouds of electron hole pairs. It also takes
into account the effect of the time dependent electrical potential generated by these coherent electron/hole pairs:
this potential acts as a quantum phase on the electrons.
Due to time translation invariance of the dynamics, it is relevant to decompose
$t_\pm=\bar{t}\pm \bar{\tau}/2$ since in this case $\mathcal{D}_{\bar{t}+\bar{\tau}/2,\bar{t}-\bar{\tau}/2}(t,\tau)$
only depends on $\tau$, $\bar{\tau}$ and $t-\bar{t}$. The full decoherence coefficient is finally equal to:
\begin{widetext}
\begin{eqnarray}
 \mathcal{D}_{\bar{t}+\bar{\tau}/2,\bar{t}-\bar{\tau}/2}(t,\tau)  
 & = & \exp{\left(\int_0^{+\infty}(1-t^*(\omega))e^{i\omega(t-\bar{t})}
\left[e^{i\omega\tau/2}-e^{-i\omega\tau/2}\right]e^{i\omega \bar{\tau}/2}
 \frac{\mathrm{d}\omega}{\omega}\right)}\nonumber \\
 & \times & 
 \exp{\left(\int_0^{+\infty}(1-t(\omega))e^{-i\omega (t-\bar{t})}
 \left[e^{-i\omega\tau/2}-e^{i\omega\tau/2}\right]e^{i\omega \bar{\tau}/2}
\frac{\mathrm{d}\omega}{\omega}\right)}\,.
\label{eq:coherence:outcoming:D} 
\end{eqnarray}
\end{widetext}
In the case of short range interactions, Eqs.~\eqref{eq:coherence:outcoming:4pt}
and \eqref{eq:coherence:outcoming:D} lead to an expression for the outcoming
single electron coherence
involving algebraic functions whose infrared divergences are not easy to control~\cite{Wahl:2013-1}.
In order to evaluate the outcoming single electron coherence, it is then more convenient
to go to the frequency domain and to use Wick's theorem to decompose the four point
correlator in Eq.~\eqref{eq:coherence:outcoming:4pt} as a sum of products of two
point correlators. Moreover, this method enables treating various effective interactions. It opens the
way to the comparison of their effect on electronic decoherence as will be discussed in a forthcoming publication.

\medskip

The outcoming single
electron coherence is thus obtained as a sum of two contributions which generalize the decomposition 
given by Eqs.~\eqref{eq:coherence:free:decomposition} and \eqref{eq:coherence:free:wp}
in the case where interactions are present:
\begin{widetext}
\begin{subequations}
\begin{eqnarray}
\mathcal{G}_{\mathrm{out}}^{(e)}\left(t+\frac{\tau}{2},t-\frac{\tau}{2}\right) & = &
\mathcal{G}_{\mathrm{out},\mathrm{mv}}^{(e)}\left(t+\frac{\tau}{2},t-\frac{\tau}{2}\right) +
\mathcal{G}_{\mathrm{out},\mathrm{wp}}^{(e)}\left(t+\frac{\tau}{2},t-\frac{\tau}{2}\right) \\
\mathcal{G}_{\mathrm{out},\mathrm{mv}}^{(e)}\left(t+\frac{\tau}{2},t-\frac{\tau}{2}\right)  & = & 
\int \varphi_e(t_+)\varphi_e(t_-)^*
K_{\mathrm{mv}}(t,\tau|t_+,t_-)\,
\,\mathrm{d}t_+\mathrm{d}t_-\\
\mathcal{G}_{\mathrm{out},\mathrm{wp}}^{(e)}\left(t+\frac{\tau}{2},t-\frac{\tau}{2}\right)  & = & 
\int \varphi_e(t_+)\varphi_e(t_-)^*
K_{\mathrm{wp}}(t,\tau|t_+,t_-)\,
\,\mathrm{d}t_+\mathrm{d}t_-\,.
\end{eqnarray}
\end{subequations}
\end{widetext}
The two propagators $K_{\mathrm{mv}}$ and $K_{\mathrm{wp}}$ are given in the time domain by:
\begin{widetext}
\begin{subequations}
\begin{eqnarray}
\label{eq:propagator:mv:time-domain}
K_{\mathrm{mv}}(t,\tau|t_+,t_-) & = & v_F^2
\left\langle \psi^\dagger\left(t-\frac{\tau}{2}\right)\psi\left(t+\frac{\tau}{2}\right)\right\rangle_F\,
\langle \psi(t_-)\psi^\dagger(t_+)\rangle_F\,
\mathcal{D}_{t_+,t_-}(t,\tau)
\\
\label{eq:propagator:wp:time-domain}
K_{\mathrm{wp}}(t,\tau|t_+,t_-) & = & v_F^2
\left\langle \psi\left(t+\frac{\tau}{2}\right)\psi^\dagger(t_+)\right\rangle_F
\left\langle \psi^\dagger\left(t-\frac{\tau}{2}\right)\psi(t_-)\right\rangle_F^*\,
\mathcal{D}_{t_+,t_-}(t,\tau)
\end{eqnarray}
\end{subequations}
\end{widetext}
In the case of weak interactions with limited bandwidth, the two
contributions to single electron coherence are well separated~\cite{Degio:2009-1}:
Eq.~\eqref{eq:propagator:mv:time-domain} contains the Fermi sea agitated by the electron/hole pairs
generated by the incoming electronic excitation through effective Coulomb interactions
whereas Eq.~\eqref{eq:propagator:wp:time-domain} encodes the contribution of the incoming single electron excitation
experiencing relaxation. 

\medskip

Due to time translation invariance, the propagator
expressed in the time domain only depends on $\tau$, $\bar{\tau}$ and
the time difference $t-\bar{t}$. Consequently, in the frequency domain, we
have:
\begin{widetext}
\begin{subequations}
\label{eq_def_propag}
 \begin{align}
 \label{eq_def_propag:mv}
 \mathcal{G}^{(e)}_{\text{mv}}
	\left(
		\omega+\frac{\Omega}{2},
		\omega-\frac{\Omega}{2}
	\right)
	& = 
	\int_{-\infty}^{+\infty} 
	\widetilde{\varphi}_e
	\left(
		\omega'+\frac{\Omega}{2}
	\right)
	\widetilde{\varphi}_e^*	
	\left(
		\omega'-\frac{\Omega}{2}
	\right)
	K_{\text{mv}}\left(\omega,\omega';\Omega\right)\,\mathrm{d}\omega'\\
\label{eq_def_propag:wp}
 \mathcal{G}^{(e)}_{\text{wp}}
	\left(
		\omega+\frac{\Omega}{2},
		\omega-\frac{\Omega}{2}
	\right)
	& = 
	\int_{-\infty}^{+\infty} 
	\widetilde{\varphi}_e
	\left(
		\omega'+\frac{\Omega}{2}
	\right)
	\widetilde{\varphi}_e^*	
	\left(
		\omega'-\frac{\Omega}{2}
	\right)
	K_{\text{wp}}\left(\omega,\omega';\Omega\right)\,\mathrm{d}\omega'  
 \end{align}
\end{subequations}
\end{widetext}
where $K_{\mathrm{mv}}(\omega,\omega';\Omega)$ and
$K_{\mathrm{wp}}(\omega,\omega';\Omega)$ are the Fourier
transforms of the two propagators defined by
Eqs.~\eqref{eq:propagator:mv:time-domain} and \eqref{eq:propagator:wp:time-domain} with respect to
$\tau$, $\bar{\tau}$ and $t-\bar{t}$. The complete propagator that appears in Eq.~(2)
of the paper is then $K(\omega,\omega';\Omega)=K_{\text{mv}}\left(\omega,\omega';\Omega\right)
+K_{\text{wp}}\left(\omega,\omega';\Omega\right)$.
We shall now derive their explicit expressions
which have been used to obtain the numerical results presented in the paper.

\subsection{Explicit expressions}

\subsubsection{The free propagators}

To begin with, for vanishing interactions, $\mathcal{D}_{t_+,t_-}(t,\tau)=1$.
In this case, the coherence propagators can easily be evaluated:
\begin{subequations}
\begin{eqnarray}
\label{eq:propagator:free:mv}
K_{\text{mv}}^{(0)}(\omega,\omega';\Omega) & = & 2\pi \delta (\Omega)\,
f_F(\omega)\,(1-f_F(\omega'))\\
K_{\text{wp}}^{(0)}(\omega,\omega';\Omega) & = & 2\pi \delta(\omega-\omega')
 \,\left(1-f_F\left(\omega+\frac{\Omega}{2}\right)\right)\nonumber \\
& \times & \left(1-f_F\left(\omega-\frac{\Omega}{2}\right)\right)\,
\label{eq:propagator:free:wp}
\end{eqnarray}
\end{subequations}
in which $f_F(\omega)=\Theta(-\omega)$ denotes the zero temperature Fermi function.
The $\delta(\Omega)$ in Eq.~\eqref{eq:propagator:free:mv} reflects the stationarity of $K^{(0)}_{\mathrm{mv}}$ which
at zero temperature, restores the Fermi sea contribution to single electron coherence. 
The $\delta(\omega-\omega')$ in Eq.~\eqref{eq:propagator:free:wp} reflects the fact that, for vanishing interactions, no dissipative
processes are present. Note that at zero temperature, $K_{\mathrm{wp}}^{(0)}$ is non vanishing 
only when both $\omega_+=\omega+\Omega/2$ and $\omega_-=\omega-\Omega/2$ are positive. 
In this case, $(\omega+\Omega/2,\omega-\Omega/2)$ is said to belong to
the (e) quadrant~\cite{Ferraro:2013-1}.
This contribution gives the excess single electron coherence associated with the incoming wave packet. Of course, we are interested
in the excess single electron coherence and therefore, we shall always subtract 
the contribution $K_{\text{mv}}^{(0)}$ that leads to the Fermi sea from the propagator.

\subsubsection{The interacting case: general properties of the propagators}

\begin{figure}[h!]
    \centering
\begin{tikzpicture}[scale=0.7]
\def\wp{0.7}
\pgfmathsetmacro\str{0.4*\wp}
\def\xlen{4}
\def\ymax{3}
\def\ymin{-3}
\def\we{1.8}
\tikzstyle{dot}=[draw,circle,minimum size=3pt,inner sep=0pt,
	outer sep=0pt,fill]

\fill[black!10] (\wp,\wp+\we) .. controls (\str,\str+\we) and (-\str,\str+\we) ..
	(-\wp,\wp+\we) --
	(-\wp,-\wp-\we) .. controls (-\str,-\str-\we) and (\str,-\str-\we) ..
	(\wp,-\wp-\we) -- cycle;

\draw[shift={(0,\we)},fill=black!20] (\wp,\wp) .. controls (\str,\str) and (\str,-\str) ..
      (\wp,-\wp) .. controls (\str,-\str) and (-\str,-\str) ..
	  (-\wp,-\wp) .. controls (-\str,-\str) and (-\str,\str) ..
	  (-\wp,\wp) .. controls (-\str,\str) and (\str,\str) ..
	  (\wp,\wp);

\draw[->] (-\xlen,0) -- (\xlen,0) node[below] {$\frac{\Omega}{2}$};
\draw[->] (0,\ymin) -- (0,\ymax) node[left] {$\omega$};

\draw[->] (\ymin,\ymin) -- (\ymax,\ymax) node[right] {$\omega_+$};
\draw[->] (-\ymin,\ymin) -- (-\ymax,\ymax) node[left] {$\omega_-$};

\pgfmathsetmacro\propagy{\we-0.25}
\def\propagx{0.28}
\coordinate[dot] () at (\propagx,\propagy);
\draw[thick] (\propagx,\propagy) -- (\propagx,-\propagy);
\draw (-2pt,\propagy) -- (2pt,\propagy) node[at start,right=10pt] {\footnotesize{$\omega'$}};
\draw (-2pt,-\propagy) -- (2pt,-\propagy) node[at start,right=6pt] {\footnotesize{$-\omega'$}};

\node[right] at (1,2.5) {(e)};
\node[right] at (1,-2.5) {(h)};
\node[right] at (2,1) {(e/h)};
\node[left] at (-2,1) {(e/h)};
\end{tikzpicture}
    \caption{%
        Propagation of single electron coherence in the frequency domain: the 
        coherence of the incoming electronic wavepacket  at $(\omega'+\Omega/2,\omega'-\Omega/2)$ 
        ($\omega'>0$) contributes to the excess outcoming coherence at 
        $(\omega+\Omega/2,\omega-\Omega/2)$ for $-\omega'\leq\omega\leq \omega'$
        (points belonging to the vertical black line).
        Grey zone: localization of the incoming
        single electron coherence in the frequency domain for a Landau quasi-particle. 
        Light grey zone: localization of the excess outcoming single electron coherence. The frequency domain
        is divided in four quadrants~\cite{Ferraro:2013-1}: two of them corresponding respectively to electron (e) and hole (h) excitations
        and two off-diagonal quadrants associated with electron/hole (e/h) coherences. Interactions
        are expected to generate both of them from an incoming excitation located within the (e) quadrant.
    }
    \label{fig_propagation_coherence}
\end{figure}
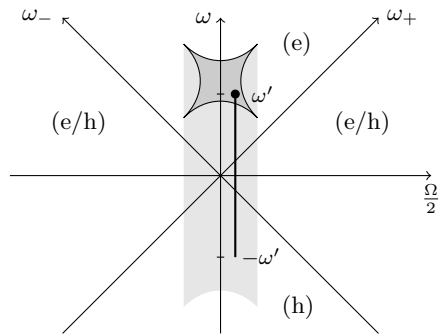

In the interacting case, the decoherence coefficient $\mathcal{D}_{t_+,t_-}(t,\tau)$ is no longer equal to unity
and should therefore be taken into account. 
Fig.~\ref{fig_propagation_coherence} depicts how the coherence
of the incoming wave packet contributes to the outcoming excess coherence after the interaction region. 

\medskip

The initial coherence only contributes to the outcoming coherence at lower energies because we
work at zero temperature and therefore
there are no heating effects. Moreover, the incoming excess coherence at 
$(\omega'+\Omega/2,\omega'-\Omega/2)$ cannot contribute to the outcoming coherence at
$(\omega+\Omega/2,\omega-\Omega/2)$ for $\omega<- \omega'$ since the incoming
electron at energy $\hbar \omega'>0$ cannot generate an electron/hole pair excitation of
energy higher than $\hbar \omega'$. The Pauli exclusion principle then leads to the
selection rule $K(\omega,\omega';\Omega)=0$ for $\omega<-\omega'$.

\subsubsection{Basic blocks}

Apart from the case of free propagation, it is impossible to find
closed compact expressions for these propagators. Consequently, a numerical 
evaluation has to be performed which, as we shall see now, can be quite challenging and
requires a careful methodology to be both effective and accurate. 

\medskip

As explained before, the best
strategy is to evaluate the propagators in the frequency domain.
In the interacting case, the novelty comes from the decoherence
coefficient $\mathcal{D}_{\bar{\tau}/2,-\bar{\tau}/2}(t+\tau/2,t-\tau/2)$. Its Fourier transform
with respect to $\tau$, $\bar{\tau}$ and $t$ can be obtained as a convolution involving the
Fourier transforms:
\begin{equation}
\Gamma_\pm(\omega)=\int_{-\infty}^{+\infty}
e^{i\omega t} \exp{\left( \pm\int_0^{+\infty} (1-t(\omega'))(e^{i\omega' t}-1)\frac{\mathrm{d}\omega'}{\omega'}
\right)}\,\mathrm{d}t\,.
\end{equation}
Assuming that 
$$\Lambda_\pm=\exp{\left(\mp \int_0^{+\infty} (1-t(\omega))\frac{\mathrm{d}\omega}{\omega}\right)}$$
is finite and non zero, which can always be ensured by a suitable UV regularization since it is known
that $t(\omega)-1\simeq O(\omega)$ at $\omega\rightarrow 0^+$, 
these Fourier transforms have a $\delta$ singularity at $\omega=0$ and a regular part
for $\omega<0$:
\begin{equation}
\label{eq:Gamma-decomposition}
	\Gamma_\pm(\omega) = 2\pi\Lambda_\pm\left( \delta(\omega) + B_\pm(-\omega)\right)
\end{equation}
where $B_\pm(\omega)$ are regular functions vanishing for $\omega< 0$. 
As we shall see, once the functions $B_\pm$ are known,
the Fourier transform of the decoherence coefficient is known.

\medskip

More precisely, using the decomposition given by Eq.~\eqref{eq:Gamma-decomposition}, 
the Fourier transform of the decoherence coefficient $\mathcal{D}_{\bar{\tau}/2,-\bar{\tau}/2}(t,\tau)$
with respect to $\tau$, $\bar{\tau}$ and $t$ can be decomposed as
a sum of $2^4=16$ terms. Therefore, each of the propagators is also a sum of 16 terms.
One among the 16 terms of the modified vacuum propagator contains the Fermi sea contribution. 
Therefore, the analytic expressions 
for the full propagator giving the excess outcoming single electron coherence
involve 31 terms which will be detailed in the present section. Although quite tedious,
these expressions provide a perfect control of all the infrared singularities arising 
in a direct time domain computation at vanishing temperature~\cite{Wahl:2013-1}. A generalization of these
expressions at non-zero temperatures can be obtained but this would go beyond the scope of
the present paper.

\medskip

Before detailing these 31 terms, let us mention that
both auxiliary functions $B_{\pm}(\omega)$ are evaluated for positive frequencies 
by solving numerically the following integral equations:
\begin{equation}
\label{eq:integro-differential}
\omega B_{\pm}(\omega) = \pm \left[1-t(\omega) + \int_0^\omega
\text{d}\omega' B_{\pm}(\omega') \left(1-t(\omega-\omega')\right) \right]
\end{equation}
where the initial value $B_{\pm}(0^+)$ is defined using the derivatives of $t(\omega)$ :
\begin{equation}
\label{eq:integro-differential:initial}
B_{\pm}(0^{+}) = \pm\lim_{\omega \to 0^{+}} \frac{1-t(\omega)}{\omega}
	= \mp t'(\omega=0^+)
\end{equation}
Expressions in the case
of effective short range interactions in the $\nu=2$ edge channel system are given in appendix~\ref{sec:explicit}. 
They have been used to check the validity of the numerical solutions for $B_\pm$ in this case. 

\subsubsection{The wave packet contribution}

The wave packet contribution is a sum of two contributions of the following form:
\begin{widetext}
\begin{subequations}
\begin{align}
	\Delta\mathcal{G}^{(e)}_{\text{wp}}
	\left(
		\omega+\frac{\Omega}{2},
		\omega-\frac{\Omega}{2}
	\right)
	&= 
	\widetilde{\varphi}_e
	\left(
		\omega+\frac{\Omega}{2}
	\right)
	\widetilde{\varphi}_e^*	
	\left(
		\omega-\frac{\Omega}{2}
	\right)
	\mathcal{Z}
	\left(
		\omega+\frac{\Omega}{2}
	\right)
	\mathcal{Z}^*
	\left(
		\omega-\frac{\Omega}{2}
	\right) \label{elasticterm}
	\\
	&+
	\int_{-\infty}^{+\infty}
		\widetilde{\varphi}_e
	\left(
		\omega'+\frac{\Omega}{2}
	\right)
	\widetilde{\varphi}_e^*	
	\left(
		\omega'-\frac{\Omega}{2}
	\right)
	K_{\text{wp}}^{\text{(ne)}}\left(\omega,\omega';\Omega\right)\, \mathrm{d}\omega'
\end{align}
\end{subequations}
\end{widetext}
The first contribution Eq.~\eqref{elasticterm} contains the purely elastic contribution corresponding to the
electronic excitation going through the interaction region without experiencing any inelastic process. 
The elastic scattering amplitude $\mathcal{Z}(\omega_{\mathrm{e}})$ for 
an electron at incoming energy $\hbar\omega_{\mathrm{e}}>0$ is
given by~\cite{Degio:2009-1}:
\begin{equation}
\mathcal{Z}(\omega_{\mathrm{e}}) = 1+\int_0^{\omega_{\mathrm{e}}} B_-(\omega')
\,\text{d}\omega'\,.
\end{equation}
Let us recall that the inelastic scattering probability for the electron at initial 
energy $\hbar\omega_{\mathrm{e}}$ is then given by
\begin{equation}
\sigma_{\mathrm{in}}(\omega_{\mathrm{e}})=1-|\mathcal{Z}(\omega_{\mathrm{e}})|^2\,.
\end{equation}
The second contribution, which contains the inelastic wave packet part 
$K_{\text{wp}}^{\text{(ne)}}$ of the propagator, is given by:
\begin{widetext}
\begin{subequations}
\begin{align}
	K_{\text{wp}}^{\text{(ne)}}\left(\omega,\omega';\Omega\right) &=
	B_+(\omega'-\omega)
	\mathcal{Z}
	\left(
		\omega+\frac{\Omega}{2}
	\right)
	\mathcal{Z}^*
	\left(
		\omega'-\frac{\Omega}{2}
	\right) + \text{ conj.}
	\\
	&+ \int_{-\infty}^{+\infty} 
	B_+(k)
	B_+^*(\omega'-\omega-k)
	\mathcal{Z}
	\left(
		\omega'+\frac{\Omega}{2}-k
	\right)
	\mathcal{Z}^*
	\left(
		\omega-\frac{\Omega}{2}+k
	\right)\,\mathrm{d} k\,
\end{align}
\end{subequations}
\end{widetext}
where the notation $f(\Omega) + \text{ conj.}$ should be understood as $f(\Omega)
+ f^*(-\Omega)$.

\subsubsection{The modified vacuum contribution}

The vacuum contribution to the excess single electron coherence
$K_{\text{mv}}$ is given by:
\begin{align}
	\Delta\mathcal{G}^{(e)}_{\text{mv}}
	\left(
		\omega+\frac{\Omega}{2},
		\omega - \frac{\Omega}{2}
	\right)
	 = 
	\int_{-\infty}^{+\infty} K_{\text{mv}} (\omega,\omega';\Omega)\, \nonumber\\
        	\times \widetilde{\varphi}_e \left(\omega'+\frac{\Omega}{2}\right)
	\widetilde{\varphi}_e^*\left(\omega'-\frac{\Omega}{2}\right)
	\text{d} \omega'\,.
\end{align}
The modified vacuum propagator $K_{\text{mv}} (\omega,\omega',\Omega)$ can
be expressed as a sum of a part arising from singularities and a part which involves no $\delta$
which is called the regular part:
\begin{equation}
	K_{\text{mv}}(\omega,\omega';\Omega) =
	K_\text{mv}^{\text{(sing)}}(\omega,\omega';\Omega) +
	\int_{\omega}^{+\infty}
	\mathcal{F}^{\text{(reg)}}_{\text{mv}}(k,\omega';\Omega)\,\mathrm{d}k
	\label{optimization}
\end{equation}
The singular part  is given by
\begin{widetext}
\begin{subequations}
\begin{align}
	K_{\text{mv}}^{\text{(sing)}}(\omega,\omega';\Omega)
	&= B_-(\Omega)\,
	\Theta\left(-\omega+\frac{\Omega}{2}\right)
	+ \text{ conj.}
	\\
	&+ B_+(\Omega)\,
	\Theta\left(-\omega-\frac{\Omega}{2}\right)
	+ \text{ conj.}
	\\
	&+\Theta\left(-\omega+\frac{\Omega}{2}\right) \int_0^{\infty} 
	B_-\left(\omega'-k+\frac{\Omega}{2}\right) B_+^*\left(\omega'-k-\frac{\Omega}{2}\right)
	\,\mathrm{d}k 
	+ \text{ conj.}\,.
\end{align}
\end{subequations}
\end{widetext}
The various Heaviside functions $\Theta$ shows that $K_{\text{mv}}^{\text{(sing)}}(\omega,\omega';\Omega)$ 
corresponds to discontinuities at the boundaries
between the (e) and (h) quadrants as well as at the boundaries
between the (e/h) and (h) quadrants~\cite{Ferraro:2013-1}. 
Finally, the regular part is the integral over $k$ from $\omega$
to infinity of:
\begin{widetext}
\begin{subequations}
\begin{align}
	\mathcal{F}_\text{mv}^{\text{(reg)}}(k,\omega';\Omega)
	&= B_+\left(-k+\frac{\Omega}{2}\right)
	B_-\left(k+\frac{\Omega}{2}\right)
	+ \text{ conj.}
	\\
	&+ B_-\left(k+\frac{\Omega}{2}\right)
	B_-^*\left(k-\frac{\Omega}{2}\right)
	\Theta(\omega'-k)
	\\
	&+ B_+\left(-k+\frac{\Omega}{2}\right)
	B_+^*\left(-k-\frac{\Omega}{2}\right)
	\Theta(\omega'+k)
	\\
	&+
	\int_{0}^{\infty} 
	B_+^*\left(\omega'-q-\frac{\Omega}{2}\right)
	B_-\left(\omega'-q+k\right)
	B_+\left(-k+\frac{\Omega}{2}\right)\,\text{d} q \,
	+ \text{ conj.}
	\\
	&+
	\int_{0}^{\infty} 
	B_-^*\left(\omega'-q-\frac{\Omega}{2}\right)
	B_+\left(\omega'-q-k\right)
	B_-\left(k+\frac{\Omega}{2}\right)\,\text{d} q
	+ \text{ conj.}
	\\
	&+
	\int_{0}^{\infty} \text{d} q
	\int_{-\infty}^{\infty}\text{d}q' \,
	B_+\left(q'-\frac{1}{2}\left(q-\omega'-\frac{\Omega}{2}+k\right)\right)
	B_-\left(-q'-\frac{1}{2}\left(k-\omega'-\frac{\Omega}{2}-k\right)\right)
	\label{mostcomplicated}
	\\
	&\times
	B_+^*\left(-q'-\frac{1}{2}\left(q-\omega'+\frac{\Omega}{2}+k\right)\right)
	B_-^*\left(q'-\frac{1}{2}\left(q-\omega'+\frac{\Omega}{2}-k\right)\right)
	\notag
\end{align}
\end{subequations}
\end{widetext}

\subsection{Numerical evaluation}
\label{appendix:numerics}

\subsubsection{Computation time}

From a numerical point of view, the wavepacket propagator $K_{\text{wp}}(\omega,\omega';\Omega)$ 
is easily evaluated since it only involves at most a single integral (the elastic scattering amplitude 
can be computed once and for all).
Assuming an isotropic discretization with $n$ points in both the $\omega$ and
$\Omega$ directions,  complexity of the wave packet contribution to
single electron coherence then scales as $\mathcal{O}(n^3)$.
The main computational difficulty arises from the term \eqref{mostcomplicated} in the
modified vacuum propagator $K_{\text{mv}}(\omega,\omega';\Omega)$. This
double integral has to be integrated twice, thus leading to a quadruple integral (complexity
$\mathcal{O}(n^4)$). Introducing a discretization on the initial coherence with $n^2$ points, 
the full evaluation would naively lead to 
a $\mathcal{O}(n^6)$ complexity for the modified vacuum contribution to the outcoming coherence.
It means that refining the discretization by a factor $2$ would increase the computation time
by a factor $64$.

\medskip

However, Eqs.~\eqref{eq_def_propag:wp} and \eqref{eq_def_propag:mv} show 
that the problem has an intrinsic anisotropy: discretization in
the $\omega$ variable should be adapted to take into account the energy scales introduced
by the dynamics itself, whereas it is reasonable to think that discretization in the $\Omega$ direction
should be determined by the time scales associated with the initial wavepacket. Therefore, introducing an anisotropic
discretization with $m$ points in the $\Omega$ direction and $n$ points in the $\omega$ direction leads to 
an $\mathcal{O}(m\times n^5)$ complexity which is still very challenging.

\medskip

\begin{figure}[h!]
	\centering
\begingroup
  \makeatletter
  \providecommand\color[2][]{%
    \GenericError{(gnuplot) \space\space\space\@spaces}{%
      Package color not loaded in conjunction with
      terminal option `colourtext'%
    }{See the gnuplot documentation for explanation.%
    }{Either use 'blacktext' in gnuplot or load the package
      color.sty in LaTeX.}%
    \renewcommand\color[2][]{}%
  }%
  \providecommand\includegraphics[2][]{%
    \GenericError{(gnuplot) \space\space\space\@spaces}{%
      Package graphicx or graphics not loaded%
    }{See the gnuplot documentation for explanation.%
    }{The gnuplot epslatex terminal needs graphicx.sty or graphics.sty.}%
    \renewcommand\includegraphics[2][]{}%
  }%
  \providecommand\rotatebox[2]{#2}%
  \@ifundefined{ifGPcolor}{%
    \newif\ifGPcolor
    \GPcolortrue
  }{}%
  \@ifundefined{ifGPblacktext}{%
    \newif\ifGPblacktext
    \GPblacktexttrue
  }{}%
  \let\gplgaddtomacro\g@addto@macro
  \gdef\gplbacktext{}%
  \gdef\gplfronttext{}%
  \makeatother
  \ifGPblacktext
    \def\colorrgb#1{}%
    \def\colorgray#1{}%
  \else
    \ifGPcolor
      \def\colorrgb#1{\color[rgb]{#1}}%
      \def\colorgray#1{\color[gray]{#1}}%
      \expandafter\def\csname LTw\endcsname{\color{white}}%
      \expandafter\def\csname LTb\endcsname{\color{black}}%
      \expandafter\def\csname LTa\endcsname{\color{black}}%
      \expandafter\def\csname LT0\endcsname{\color[rgb]{1,0,0}}%
      \expandafter\def\csname LT1\endcsname{\color[rgb]{0,1,0}}%
      \expandafter\def\csname LT2\endcsname{\color[rgb]{0,0,1}}%
      \expandafter\def\csname LT3\endcsname{\color[rgb]{1,0,1}}%
      \expandafter\def\csname LT4\endcsname{\color[rgb]{0,1,1}}%
      \expandafter\def\csname LT5\endcsname{\color[rgb]{1,1,0}}%
      \expandafter\def\csname LT6\endcsname{\color[rgb]{0,0,0}}%
      \expandafter\def\csname LT7\endcsname{\color[rgb]{1,0.3,0}}%
      \expandafter\def\csname LT8\endcsname{\color[rgb]{0.5,0.5,0.5}}%
    \else
      \def\colorrgb#1{\color{black}}%
      \def\colorgray#1{\color[gray]{#1}}%
      \expandafter\def\csname LTw\endcsname{\color{white}}%
      \expandafter\def\csname LTb\endcsname{\color{black}}%
      \expandafter\def\csname LTa\endcsname{\color{black}}%
      \expandafter\def\csname LT0\endcsname{\color{black}}%
      \expandafter\def\csname LT1\endcsname{\color{black}}%
      \expandafter\def\csname LT2\endcsname{\color{black}}%
      \expandafter\def\csname LT3\endcsname{\color{black}}%
      \expandafter\def\csname LT4\endcsname{\color{black}}%
      \expandafter\def\csname LT5\endcsname{\color{black}}%
      \expandafter\def\csname LT6\endcsname{\color{black}}%
      \expandafter\def\csname LT7\endcsname{\color{black}}%
      \expandafter\def\csname LT8\endcsname{\color{black}}%
    \fi
  \fi
  \setlength{\unitlength}{0.0500bp}%
  \begin{picture}(4534.00,2834.00)%
    \gplgaddtomacro\gplbacktext{%
      \csname LTb\endcsname%
      \put(1078,704){\makebox(0,0)[r]{\strut{} 0.1}}%
      \put(1078,1170){\makebox(0,0)[r]{\strut{} 1}}%
      \put(1078,1637){\makebox(0,0)[r]{\strut{} 10}}%
      \put(1078,2103){\makebox(0,0)[r]{\strut{} 100}}%
      \put(1078,2569){\makebox(0,0)[r]{\strut{} 1000}}%
      \put(1333,484){\makebox(0,0){\strut{} 100}}%
      \put(4026,484){\makebox(0,0){\strut{} 1000}}%
      \put(176,1636){\rotatebox{-270}{\makebox(0,0){\strut{} Computation time (min)}}}%
      \put(2673,154){\makebox(0,0){\strut{}$n$}}%
    }%
    \gplgaddtomacro\gplfronttext{%
    }%
    \gplbacktext
    \put(0,0){\includegraphics{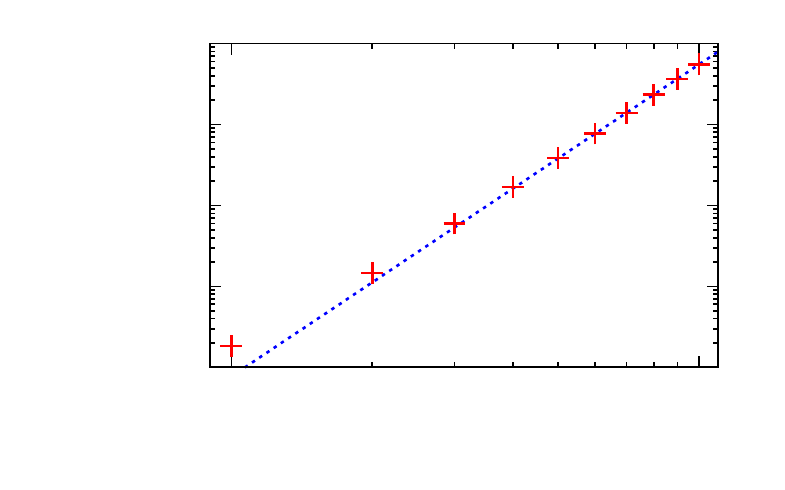}}%
    \gplfronttext
  \end{picture}%
\endgroup
	\caption{%
	Computation time as a function of the number $n$ of points in the $\omega$ direction. Dashed
	curve is a linear fit in a log-log plot. Best fit is obtained with $t =\mathcal{O}(n^{3.85})$.
	}
	\label{fig:complexity}
\end{figure}

Finally, Eq.~\eqref{optimization} can be used to decrease the complexity by
one order of magnitude. At fixed, $\Omega$, $K_{\text{mv}}$ can be computed
by decreasing $\omega$ and storing the integral of $\mathcal{F}_{\text{mv}}^{\text{(reg)}}$.
Consequently only one evaluation of $\mathcal{F}_{\text{mv}}^{\text{(reg)}}$ is required
and the total complexity can then be brought down to $\mathcal{O}(m\times n^4)$. Fortunately,
the problem is well suited to parallelization: our C code exploits the OpenMP API interface and
runs on a Dell PowerEdge R815 computer with 30 Opteron cores. Fig.~\ref{fig:complexity}
shows the scaling of the computing time as a
function of the number of points $n$ in the $\omega$ direction: a doubling $n\mapsto 2n$ leads 
to an increase by a factor $\sim 15$, close to the expected $2^4=16$.

\subsubsection{Testing the accuracy of the code}

We have devised specific indicators to test the accuracy of the code: they
are based on sum rules such as
total charge conservation which had been used in our study of the relaxation of an 
electronic excitation with perfectly defined energy~\cite{Degio:2009-1}. In time
dependent situations considered in the present paper, this sum rule is generalized into sum rules
involving the time dependent average electric current.

\medskip

\begin{figure*}
	\centering
\begingroup
  \makeatletter
  \providecommand\color[2][]{%
    \GenericError{(gnuplot) \space\space\space\@spaces}{%
      Package color not loaded in conjunction with
      terminal option `colourtext'%
    }{See the gnuplot documentation for explanation.%
    }{Either use 'blacktext' in gnuplot or load the package
      color.sty in LaTeX.}%
    \renewcommand\color[2][]{}%
  }%
  \providecommand\includegraphics[2][]{%
    \GenericError{(gnuplot) \space\space\space\@spaces}{%
      Package graphicx or graphics not loaded%
    }{See the gnuplot documentation for explanation.%
    }{The gnuplot epslatex terminal needs graphicx.sty or graphics.sty.}%
    \renewcommand\includegraphics[2][]{}%
  }%
  \providecommand\rotatebox[2]{#2}%
  \@ifundefined{ifGPcolor}{%
    \newif\ifGPcolor
    \GPcolorfalse
  }{}%
  \@ifundefined{ifGPblacktext}{%
    \newif\ifGPblacktext
    \GPblacktexttrue
  }{}%
  \let\gplgaddtomacro\g@addto@macro
  \gdef\gplbacktext{}%
  \gdef\gplfronttext{}%
  \makeatother
  \ifGPblacktext
    \def\colorrgb#1{}%
    \def\colorgray#1{}%
  \else
    \ifGPcolor
      \def\colorrgb#1{\color[rgb]{#1}}%
      \def\colorgray#1{\color[gray]{#1}}%
      \expandafter\def\csname LTw\endcsname{\color{white}}%
      \expandafter\def\csname LTb\endcsname{\color{black}}%
      \expandafter\def\csname LTa\endcsname{\color{black}}%
      \expandafter\def\csname LT0\endcsname{\color[rgb]{1,0,0}}%
      \expandafter\def\csname LT1\endcsname{\color[rgb]{0,1,0}}%
      \expandafter\def\csname LT2\endcsname{\color[rgb]{0,0,1}}%
      \expandafter\def\csname LT3\endcsname{\color[rgb]{1,0,1}}%
      \expandafter\def\csname LT4\endcsname{\color[rgb]{0,1,1}}%
      \expandafter\def\csname LT5\endcsname{\color[rgb]{1,1,0}}%
      \expandafter\def\csname LT6\endcsname{\color[rgb]{0,0,0}}%
      \expandafter\def\csname LT7\endcsname{\color[rgb]{1,0.3,0}}%
      \expandafter\def\csname LT8\endcsname{\color[rgb]{0.5,0.5,0.5}}%
    \else
      \def\colorrgb#1{\color{black}}%
      \def\colorgray#1{\color[gray]{#1}}%
      \expandafter\def\csname LTw\endcsname{\color{white}}%
      \expandafter\def\csname LTb\endcsname{\color{black}}%
      \expandafter\def\csname LTa\endcsname{\color{black}}%
      \expandafter\def\csname LT0\endcsname{\color{black}}%
      \expandafter\def\csname LT1\endcsname{\color{black}}%
      \expandafter\def\csname LT2\endcsname{\color{black}}%
      \expandafter\def\csname LT3\endcsname{\color{black}}%
      \expandafter\def\csname LT4\endcsname{\color{black}}%
      \expandafter\def\csname LT5\endcsname{\color{black}}%
      \expandafter\def\csname LT6\endcsname{\color{black}}%
      \expandafter\def\csname LT7\endcsname{\color{black}}%
      \expandafter\def\csname LT8\endcsname{\color{black}}%
    \fi
  \fi
  \setlength{\unitlength}{0.0500bp}%
  \begin{picture}(9070.00,3400.00)%
    \gplgaddtomacro\gplbacktext{%
      \csname LTb\endcsname%
      \put(1405,3390){\makebox(0,0){\strut{}(a) $\tau = 0$}}%
    }%
    \gplgaddtomacro\gplfronttext{%
      \csname LTb\endcsname%
      \put(765,395){\makebox(0,0){\strut{}-2}}%
      \put(1085,395){\makebox(0,0){\strut{} 0}}%
      \put(1405,395){\makebox(0,0){\strut{} 2}}%
      \put(1725,395){\makebox(0,0){\strut{} 4}}%
      \put(2045,395){\makebox(0,0){\strut{} 6}}%
      \put(1405,65){\makebox(0,0){\strut{}$\bar{t}/\tau_e$}}%
      \put(433,681){\makebox(0,0)[r]{\strut{}-5}}%
      \put(433,1157){\makebox(0,0)[r]{\strut{} 0}}%
      \put(433,1633){\makebox(0,0)[r]{\strut{} 5}}%
      \put(433,2107){\makebox(0,0)[r]{\strut{} 10}}%
      \put(433,2583){\makebox(0,0)[r]{\strut{} 15}}%
      \put(433,3059){\makebox(0,0)[r]{\strut{} 20}}%
      \put(-29,1870){\rotatebox{-270}{\makebox(0,0){\strut{}$\overline{\Omega}\tau_{\mathrm{e}}$}}}%
    }%
    \gplgaddtomacro\gplbacktext{%
      \csname LTb\endcsname%
      \put(4171,3390){\makebox(0,0){\strut{}(b) $\tau = 0.4\tau_{\mathrm{e}}$}}%
    }%
    \gplgaddtomacro\gplfronttext{%
      \csname LTb\endcsname%
      \put(3557,395){\makebox(0,0){\strut{}-2}}%
      \put(3864,395){\makebox(0,0){\strut{} 0}}%
      \put(4171,395){\makebox(0,0){\strut{} 2}}%
      \put(4478,395){\makebox(0,0){\strut{} 4}}%
      \put(4785,395){\makebox(0,0){\strut{} 6}}%
      \put(4171,65){\makebox(0,0){\strut{}$\bar{t}/\tau_e$}}%
      \put(3232,681){\makebox(0,0)[r]{\strut{}-5}}%
      \put(3232,1157){\makebox(0,0)[r]{\strut{} 0}}%
      \put(3232,1633){\makebox(0,0)[r]{\strut{} 5}}%
      \put(3232,2107){\makebox(0,0)[r]{\strut{} 10}}%
      \put(3232,2583){\makebox(0,0)[r]{\strut{} 15}}%
      \put(3232,3059){\makebox(0,0)[r]{\strut{} 20}}%
      \put(2770,1870){\rotatebox{-270}{\makebox(0,0){\strut{}$\omega\tau_{\mathrm{e}}$}}}%
    }%
    \gplgaddtomacro\gplbacktext{%
      \csname LTb\endcsname%
      \put(6983,3390){\makebox(0,0){\strut{}(c) $\tau = 0.8\tau_e$}}%
    }%
    \gplgaddtomacro\gplfronttext{%
      \csname LTb\endcsname%
      \put(6369,395){\makebox(0,0){\strut{}-2}}%
      \put(6676,395){\makebox(0,0){\strut{} 0}}%
      \put(6983,395){\makebox(0,0){\strut{} 2}}%
      \put(7290,395){\makebox(0,0){\strut{} 4}}%
      \put(7597,395){\makebox(0,0){\strut{} 6}}%
      \put(6983,65){\makebox(0,0){\strut{}$\bar{t}/\tau_e$}}%
      \put(6044,681){\makebox(0,0)[r]{\strut{}-5}}%
      \put(6044,1157){\makebox(0,0)[r]{\strut{} 0}}%
      \put(6044,1633){\makebox(0,0)[r]{\strut{} 5}}%
      \put(6044,2107){\makebox(0,0)[r]{\strut{} 10}}%
      \put(6044,2583){\makebox(0,0)[r]{\strut{} 15}}%
      \put(6044,3059){\makebox(0,0)[r]{\strut{} 20}}%
      \put(5582,1870){\rotatebox{-270}{\makebox(0,0){\strut{}$\omega\tau_{\mathrm{e}}$}}}%
      \put(8793,1155){\makebox(0,0)[l]{\strut{} 0}}%
      \put(8793,1750){\makebox(0,0)[l]{\strut{} 0.5}}%
      \put(8793,2345){\makebox(0,0)[l]{\strut{} 1}}%
      \put(8793,2940){\makebox(0,0)[l]{\strut{} 1.5}}%
    }%
    \gplbacktext
    \put(0,0){\includegraphics[width=16cm]{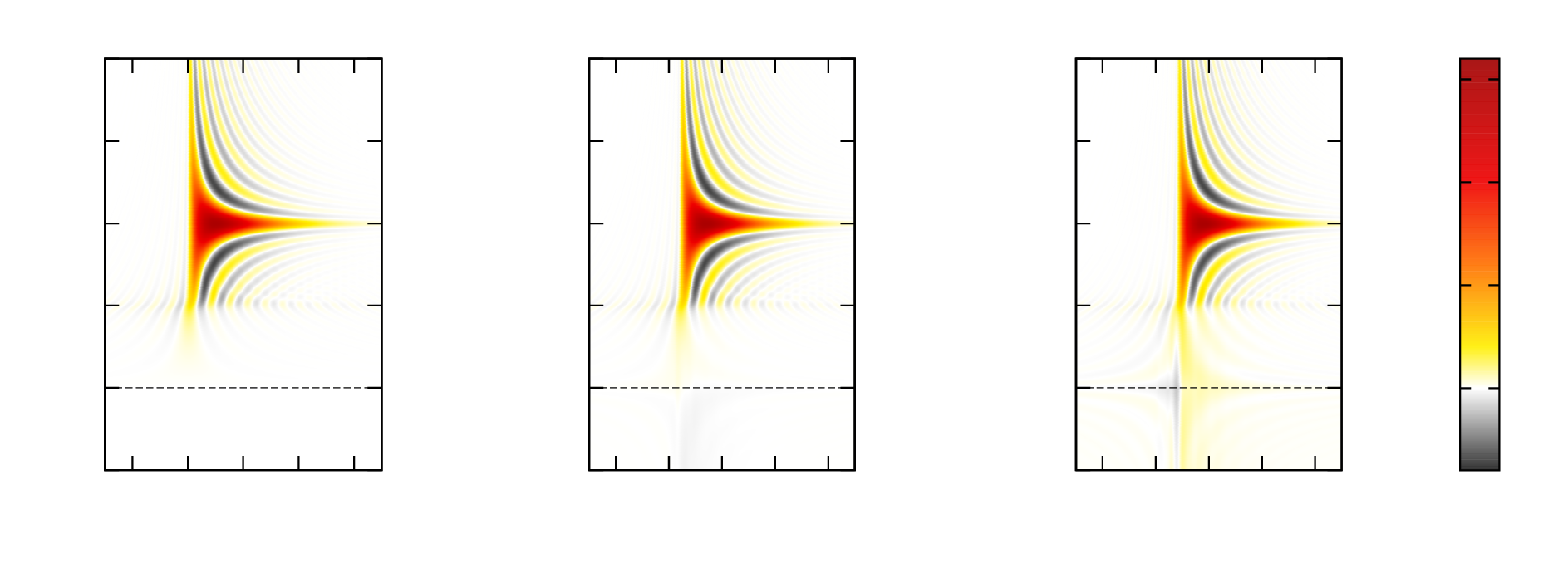}}%
    \gplfronttext
  \end{picture}%
\endgroup
	\caption{\label{fig:errors}
		Excess Wigner function $\Delta W^{(e)}(t,\omega)$ after free propagation for various time of flight
		$\tau$: (a) incoming excitation with $\omega_{\mathrm{e}}\tau_{\mathrm{e}}=10$,
		(b) outcoming excess Wigner function for $\tau=0.4\tau_{\mathrm{e}}$ and
		(c) for $\tau=0.8\tau_{\mathrm{e}}$. In the last case, numerical errors are clearly visible. These figures were
		obtained using a discretization with $n=1024$ points in the $\omega$ direction and $m=5\times 1024$
		in the $\Omega$ direction (time resolution $0.052$ in $t/\tau_{\mathrm{e}}$).
	}
\end{figure*}

More precisely, due to the relation between the Fourier
components of the electrical current and the edge magnetoplasmon modes~\cite{Grenier:2013-1}, 
the average outcoming electric current can be computed straightforwardly from
the incoming current and the edge magnetoplasmon transmission amplitude $t(\omega)$:
\begin{subequations}
\label{eq:current:2}
\begin{eqnarray}
	\label{eq:current:2a}
	\langle i(t) \rangle & = &
	-ev_F
	\int_{-\infty}^{+\infty} \mathrm{d}\tau \,
		\chi(\tau) |\varphi_e(t-\tau)|^2
	\\
	\label{eq:current:2b}
	\chi(\tau) & = & \int_0^{\infty} \left( t(\omega) e^{-i \omega \tau} + \text{c.c.}\right)\,\frac{\mathrm{d}\omega}{2\pi}\,.
\end{eqnarray}
\end{subequations}
But it can also be computed directly from the excess Wigner function obtained by
DFT from the excess single electron coherence in the frequency domain:
\begin{equation}
	\label{eq:current:1}
	\langle i(t) \rangle =
	- e\int_{-\infty}^{+\infty} \Delta W_{\mathrm{out}}^{(e)}(t,\omega)\,
\frac{\mathrm{d} \omega}{2\pi} \, 
\end{equation}
In practice, the case of a free propagation ($t(\omega)=e^{i\omega\tau}$) provides a stringent
test of the code accuracy since increasing the time of flight $\tau$ introduces more
and more oscillations in all the integrals that are evaluated. Free propagation is the worst situation for
numerical convergence: in the case of
an incoming Landau quasiparticle of duration $\tau_{\mathrm{e}}$ and injected
at energy $\hbar\omega_{\mathrm{e}}$ such that $\omega_{\mathrm{e}}\tau_{\mathrm{e}}=10$, 
the appearance of numerical errors can be seen on Fig.~\ref{fig:errors}: at $\tau=0.8\tau_{\mathrm{e}}$,
one clearly see that the computed outcoming excess Wigner function exhibits 
hole contributions which should not be present. The total charge departs
from the expected value by 5~\% for $\tau=0.4\tau_{\mathrm{e}}$ and 
by 45~\% for $\tau=0.8\tau_{\mathrm{e}}$ and the 
electrical current \eqref{eq:current:1} computed from the excess Wigner
function departs by 8~\% from the expected one \eqref{eq:current:2a} for 
$\tau=0.4\tau_{\mathrm{e}}$ and by 25~\% for $\tau=0.8\tau_{\mathrm{e}}$.

\medskip

In the presence of interaction, the error rate is much
lower since the integrals involved in the numerical computation do not involve purely
oscillating functions. In order to avoid these problems and maximize the convergence
speed of the numerics, we have also compensated for the global drift of the 
fractionalized current pulses by introducing a phase $e^{-i\omega\bar\tau}$
where $\bar\tau=(\tau_c+\tau_s)/2$ in front of the edge magnetoplasmon scattering matrix.
Finally, the error rates corresponding to the interacting 
situations considered in the paper are much lower than the one discussed in the previous paragraph.
Using parameters given in Table \ref{table:parameters},
for Landau excitations at $\tau_s=\tau_{\mathrm{e}}$,
the total charge departs from the expected value by $1.8\ \%$ and the current
by $6.3\ \%$. For Levitov excitations at $\tau_s=20\tau_0$, the total charge departs
from the expected value by $4.9\ \%$ and the electrical current by $1.2\ \%$.

\subsubsection{Parameters choice}

Table \ref{table:parameters} summarizes the choice of parameters used to generate the results presented in the
present paper. Discretization in the $\omega$ direction has been chosen to maximize accuracy of the results
while keeping the computation time reasonable. This is challenging since a small discretization step in $\omega$
is required to ensure convergence of the numerical estimations of the
integrals due to terms which oscillate faster with increasing copropagation distance. 
Moreover a sufficiently large interval for $\Omega$ is also needed so that the time dependance
of the Wigner function over short time scales ($\lesssim \tau_{\mathrm{e}}$ or
$\lesssim \tau_{0}$) is captured. 

\medskip

The difference between the scales chosen between the Landau quasi-particle
and the Leviton are justified by the expected scales of variations of the single electron coherence: 
in the frequency space, the Leviton coherence is located close to the Fermi surface
and decays exponentially ( $\Delta\mathcal{G}^{(e)}(\omega+\Omega/2,\omega-\Omega/2)$ decays 
as $\exp{(-2\omega\tau_0)}$ for $\omega>0$) whereas the Landau quasi-particle coherence
decays algebraically. Consequently, we had to consider a much larger sampling domain for the coherence
of Landau quasi-particles than for the Leviton. 

\medskip

\begin{table}
\begin{tabular}{|l|c|c|c|c|c|}\hline
Excitation & $n$ & $m$ & $\omega_{\text{max}}/2\pi$ & $\Omega_{\text{max}}/2\pi$ & $\delta t$ \\ \hline \hline
Landau & $1024$ & $5\times 1024$ &  $3$ & $600$ & $0.0052$ \\ \hline
Leviton & $1024$ & $5\times 1024$ &  $30$ & $60$ & $0.052$ \\ \hline
\end{tabular}
\caption{\label{table:parameters} Parameters used to generate the figures presented in the paper for both 
the Landau and Leviton excitations. All parameters are given in units of the characteristic time scale
of the excitation: $\tau_{\mathrm{e}}$ for the Landau quasi-particle and $\tau_{0}$ for the Leviton.}
\end{table}

Finally, videos have been generated to illustrate the evolution of the excess Wigner function
for Landau and Levitov quasi-particles with increasing copropagation distances. Each of these
videos involves 100 images.

\section{Basic blocks for the case of $\nu=2$}
\label{sec:explicit}

In the case of the $\nu=2$ edge channel system, the auxiliary functions $B_\pm(\omega)$
can be computed analytically in closed form:
\begin{widetext}
\begin{subequations}
\begin{eqnarray}
B_+(\omega) & = & -i \tau_c^{p_c}\tau_s^{p_s}\Theta(\omega)+\frac{p_sp_c}{2}
+ \frac{(\tau_s-\tau_c)^2}{\tau_c}e^{-i\omega\tau_c}\,
\phi_1\left[\frac{1}{2}+p_c,1,3;1-\frac{\tau_s}{\tau_c},i\omega(\tau_s-\tau_c)\right] \\
B_-(\omega) & = & \Theta(\omega)\left( 
i\tau_c\,{}_1\mathrm{F}_1\left[p_c,1;i\omega(\tau_s-\tau_c)\right] 
+i(\tau_s-\tau_c)p_c\,{}_1\mathrm{F}_1\left[\frac{1}{2}+p_c,2;-i\omega(\tau_s-\tau_c)\right]
\right)
\end{eqnarray}
\end{subequations}
\end{widetext}
where $p_{c/s}=(1\pm \cos{(\theta)})/2$ and
${}_1\mathrm{F}_1(\alpha,\beta;z)$ denotes the confluent hypergeometric function and $\phi_1(\alpha,\beta,\gamma;x,y)$ denotes
the Humbert double series~\cite{Book:GradRyz}.
Finally, an analytical
expression can also be obtained for the elastic scattering amplitude in term of the confluent hypergeometric 
function ${}_1F_1$:
\begin{equation}
\label{eq:elastic-amplitude}
\mathcal{Z}(\omega_{\mathrm{e}})=e^{i\omega\tau_c}
\,{}_1\mathrm{F}_1\left[\frac{1+\cos{(\theta)}}{2},1;i\omega_{\mathrm{e}}(\tau_s-\tau_c)\right]\,.
\end{equation}
which reduces to a Bessel function at strong coupling. These analytical expressions have been used
to check the numerical solution of the integrodifferential equation \eqref{eq:integro-differential} with initial 
condition given by Eq.~\eqref{eq:integro-differential:initial}.


\end{document}